\documentclass[twocolumn]{aastex701}
\pdfoutput=1 
\usepackage[T1]{fontenc}

\usepackage{textcomp}
      
\usepackage[figure,figure*]{hypcap}
\usepackage{amsmath,amstext,bm}
\numberwithin{equation}{section}
\usepackage{graphicx,textcomp,fancyhdr}
\usepackage{tabularx}

\usepackage{float}               
\makeatletter                    
\let\newfloat\newfloat@ltx       
\makeatother                     
\usepackage{algorithm}           
\usepackage{algpseudocode}       
\makeatletter                    
\renewcommand{\fnum@algorithm}{\fname@algorithm\ \thealgorithm:}                  
\makeatother                     

\usepackage{hyperref}

\shorttitle{Integrated GPs as State-Space Models}
\shortauthors{Rubenzahl and Hattori et al.}

\DeclareRobustCommand{\okina}{%
\raisebox{\dimexpr\fontcharht\font`A-\height}{%
    \scalebox{0.8}{`}%
  }%
}

\begin{document}

\title{Scalable Gaussian Processes for Integrated and Overlapping Measurements\\ via Augmented State-Space Models}

\newcommand{\cca}{Center for Computational Astrophysics, Flatiron Institute, 162 Fifth Avenue, New York, NY 10010, USA}
\newcommand{\jpl}{Jet Propulsion Lab, Pasadena, CA 91125, USA}
\newcommand{\columbia}{Department of Astronomy, Columbia University, 538 West 120th Street, Pupin Hall, New York, NY 10027, USA}
\newcommand{\amnh}{American Museum of Natural History, 200 Central Park West, Manhattan, NY 10024, USA}
\newcommand{\stonybrook}{Department of Physics and Astronomy, Stony Brook University, Stony Brook, NY 11794, USA}
\newcommand{\nyu}{Department of Physics, New York University, 726 Broadway, New York, NY 10003, USA}
\newcommand{\google}{\red{todo: correct google affiliation}}
\newcommand{\caltechastro}{Department of Astronomy, California Institute of Technology, Pasadena, CA 91125, USA}
\newcommand{\SSL}{Space Sciences Laboratory, University of California, Berkeley, CA 94720, USA}
\newcommand{\wmko}{W. M. Keck Observatory, Waimea, HI 96743, USA}
\newcommand{\UCO}{UC Observatories, University of California, Santa Cruz, CA 95064, USA}
\newcommand{\ipac}{NASA Exoplanet Science Institute/Caltech-IPAC, MC 314-6, 1200 E. California Blvd., Pasadena, CA 91125, USA}

\author[0000-0003-3856-3143]{Ryan A. Rubenzahl}
\affiliation{\cca}
\email[show]{rrubenzahl@flatironinstitute.org}

\author[0000-0002-0842-863X]{Soichiro Hattori}
\affiliation{\columbia}
\affiliation{\amnh}
\email{soichiro.hattori@columbia.edu}

\author[0000-0002-7031-9354]{Simo S{\"a}rkk{\"a}}
\affiliation{Department of Electrical Engineering and Automation, Aalto University, 02150 Espoo, Finland}
\affiliation{ELLIS Institute Finland, Aalto University, 02150 Espoo, Finland}
\email{simo.sarkka@aalto.fi}

\author[0000-0003-1540-8562]{Will M. Farr}
\affiliation{\stonybrook}
\affiliation{\cca}
\email{wfarr@flatironinstitute.org}

\author[0000-0002-4927-9925]{Jacob K. Luhn}
\affiliation{\jpl}
\email{jacob.luhn@jpl.nasa.gov}

\author[0000-0001-9907-7742]{Megan Bedell}
\affiliation{\cca}
\email{mbedell@flatironinstitute.org}

\author[0000-0002-9328-5652]{Daniel Foreman-Mackey}
\affiliation{\cca}
\email{foreman.mackey@gmail.com}



\DeclareRobustCommand{\okina}{%
\raisebox{\dimexpr\fontcharht\font`A-\height}{%
    \scalebox{0.8}{`}%
  }%
}

\newcommand{\red}[1]{{\color{red}#1}}

\newcommand{\tess}{\textit{TESS}}
\newcommand{\kepler}{\textit{Kepler}}

\newcommand{\ms}{m~s$^{-1}$}
\newcommand{\mms}{mm~s$^{-1}$}
\newcommand{\cms}{cm~s$^{-1}$}
\newcommand{\kms}{km~s$^{-1}$}

\newcommand{\lat}{\text{lat}}
\newcommand{\lon}{\text{lon}}
\newcommand{\FWHM}{\text{FWHM}}

\newcommand{\Porb}{P_\text{orb}}
\newcommand{\istar}{i_\star}
\newcommand{\veq}{v_{eq}}
\newcommand{\sini}{\sin \istar}
\newcommand{\veqsini}{\veq\sini}
\newcommand{\avgmu}{\langle\mu\rangle}
\newcommand{\iorb}{i_\text{orb}}
\newcommand{\iorbb}{i_{\text{orb},b}}
\newcommand{\iorbc}{i_{\text{orb},c}}
\newcommand{\vsini}{v\sin\istar}
\newcommand{\Lsun}{\text{L}_\odot}
\newcommand{\Msun}{\text{M}_\odot}
\newcommand{\Rsun}{\text{R}_\odot}
\newcommand{\Mearth}{\text{M}_\oplus}
\newcommand{\Rearth}{\text{R}_\oplus}
\newcommand{\Mjup}{\text{M}_\text{Jup}}
\newcommand{\Rjup}{\text{R}_\text{Jup}}
\newcommand{\AU}{\text{AU}}
\newcommand{\Wmsq}[0]{W~m$^{-2}$}
\newcommand{\hjd}{HJD$_\text{TDB}$}
\newcommand{\bjd}{BJD$_\text{TDB}$}
\newcommand{\teff}{T_\text{eff}}
\newcommand{\Teff}{T_\text{eff}}
\newcommand{\feh}{\ensuremath{[\mbox{Fe}/\mbox{H}]}}
\newcommand{\rphk}{\ensuremath{R'_{\mbox{\scriptsize HK}}}}
\newcommand{\lrphk}{\ensuremath{\log{\rphk}}}
\newcommand{\logg}{\ensuremath{\log{g}}}

\newcommand{\bigO}[1]{$\mathcal{O}(#1)$}

\newcommand{\tabcellwidth}{0.3\textwidth}
\newcommand{\Normal}{\mathcal{N}}

\newcommand{\dd}{\mathop{}\!\mathrm{d}}

\newcommand{\shoarg}{\tau_k}
\newcommand{\sinarg}{\sin\shoarg}
\newcommand{\cosarg}{\cos\shoarg}
\newcommand{\sinharg}{\sinh\shoarg}
\newcommand{\cosharg}{\cosh\shoarg}
\newcommand{\exparg}{e^{-\frac{\omega_0\Delta}{2Q}}}
\newcommand{\expaarg}{e^{-\frac{\omega_0\Delta}{Q}}}
\newcommand{\expparg}{e^{\frac{\omega_0\Delta}{2Q}}}
\newcommand{\exppaarg}{e^{\frac{\omega_0\Delta}{Q}}}

\newcommand{\Qaa}{e^{\frac{\omega_0\Delta_k}{Q}} - 1 - \frac{1}{2\eta Q}\sin(2\shoarg) - \frac{1}{2\eta^2 Q^2} \sin^2\shoarg}
\newcommand{\Qab}{\frac{\omega_0}{\eta^2Q}\sin^2\shoarg}
\newcommand{\Qba}{\Qab}
\newcommand{\Qbb}{\omega_0^2 \left[ e^{\frac{\omega_0\Delta_k}{Q}} - 1 + \frac{1}{2\eta Q}\sin(2\shoarg) - \frac{1}{2\eta^2 Q^2} \sin^2\shoarg  \right]}

\newcommand{\Is}{a\sin\tau - b\cos\tau}
\newcommand{\Ic}{a\cos\tau + b\sin\tau}


\newcommand{\Qc}{\bm{Q}_c}
\newcommand{\m}{\bm{m}}
\renewcommand{\P}{\bm{P}}
\newcommand{\A}{\bm{A}}
\newcommand{\Q}{\bm{Q}}
\newcommand{\F}{\bm{F}}
\renewcommand{\H}{\bm{H}} 
\renewcommand{\L}{\bm{L}}
\renewcommand{\v}{\bm{v}}
\renewcommand{\S}{\bm{S}}
\newcommand{\K}{\bm{K}}
\newcommand{\G}{\bm{G}}
\newcommand{\R}{\bm{R}}
\newcommand{\I}{\bm{I}}
\newcommand{\Z}{\bm{0}}
\newcommand{\Pinf}{\bm{P}_\infty}

\newcommand{\xaug}{\tilde{\bm{x}}}
\newcommand{\Iaug}{\tilde{\bm{I}}}
\newcommand{\Haug}{\tilde{\bm{H}}}
\newcommand{\Faug}{\tilde{\bm{F}}}
\newcommand{\Laug}{\tilde{\bm{L}}}
\newcommand{\Aaug}{\tilde{\bm{A}}}
\newcommand{\Qaug}{\tilde{\bm{Q}}}
\newcommand{\Phiaug}{\tilde{\bm{\Phi}}}
\newcommand{\Phibar}{\bar{\bm{\Phi}}}
\newcommand{\RESET}{\bm{\mathcal{R}}}

\newcommand{\rdotscorner}{%
  \begin{matrix}
    \vdots \\[-0.6em]
    \smash{\raisebox{-0.2em}{\(\cdots\)}}
  \end{matrix}
}

\newcommand{\project}[1]{\textsf{#1}}
\newcommand{\celerite}{\project{celerite}}
\newcommand{\tinygp}{\project{tinygp}}
\newcommand{\smolgp}{\project{smolgp}}

\defcitealias{Sarkka2021}{SG21}
\defcitealias{LuhnIntGP}{L26}

\begin{abstract}

Astronomical measurements are often integrated over finite exposures, which can obscure latent variability on comparable timescales. Correctly accounting for exposure integration with Gaussian Processes (GPs) in such scenarios is essential but computationally challenging: once exposure times vary or overlap across measurements, the covariance matrix forfeits any quasiseparability, forcing \bigO{N^2} memory and \bigO{N^3} runtime costs. Linear Gaussian state-space models (SSMs) are equivalent to GPs and have well-known \bigO{N} solutions via the Kalman filter and Rauch--Tung--Striebel smoother. In this work, we extend the GP--SSM equivalence to handle integrated measurements while maintaining scalability by augmenting the SSM with an integral state that resets at exposure start times and is observed at exposure end times. This construction yields exactly the same posterior as a fully integrated GP but in \bigO{N} time on a CPU, and is parallelizable down to \bigO{N/T + \log T} on a GPU with $T$ parallel workers. We present State-space Models for O(Linear/log) GPs (\smolgp), an open-source Python/JAX package offering drop-in compatibility with \tinygp\ while supporting both standard and exposure-aware GP modeling. Since SSMs provide a framework for representing general GP kernels via their series expansion, \smolgp\ also brings scalable performance to many commonly used covariance kernels in astronomy that lack quasiseparability, such as the quasiperiodic kernel. The substantial performance boosts at large $N$ will enable massive multi-instrument cross-comparisons where exposure overlap is ubiquitous, and unlocks the potential for analyses with more complex models and/or higher dimensional datasets.

\end{abstract}

\section{Introduction}\label{sec:intro}

Astronomical measurements are often obtained by integrating a signal over a finite period of time, such as accumulating photoelectrons in a CCD pixel during an exposure. If the signal is time variable, the image read-out by the CCD represents the average of that latent signal over the exposure. The timescale associated with the astrophysical phenomenon being probed is usually many times longer than the length of an exposure, and so instantaneous models are effective unbiased estimators of the latent process. However, when the exposure length becomes an appreciable fraction or more of the timescale of interest, especially for stochastic signals, the effects of exposure integration must be accounted for.

Many astronomical time series of stochastic phenomena are well-modeled by a Gaussian Process \citep[GP;][]{gpml}; see \citet{AigrainDFM2023} for a review. One such example is the case of apparent radial velocity (RV) variability induced by dynamic stellar surfaces, which involve a number of distinct but related processes spanning all timescales from minutes to decades \citep{eprv_working_group_report}. The fastest of these--acoustic (e.g. \textit{p}-mode) oscillations \citep{Kjeldsen1995,Huber2011}--are around 5.5~minutes for solar-type stars, which is comparable to the length of a typical exposure. In fact, exposures are often intentionally tuned to the expected oscillation timescale \citep[from scaling relations, e.g.][]{Brown1991,Kjeldsen1995} (or an integer-multiple) in an attempt to mitigate their effect on the measured spectrum \citep{Dumusque2011a, Chaplin2019}.

There is significant utility in combining data-streams from multiple instruments to better resolve subtle signals. In the context of RV (more generally, spectral) variability, \citet{LuhnIntGP} employ GPs to isolate common astrophysical variability from individual instrumental systematics by jointly modeling the RV time series from multiple EPRV solar feeds. The most constraining information as to whether a signal is shared or not across multiple instruments comes from the times where those instruments were simultaneously observing; as such, overlapping exposures are valuable for accurate signal identification. However, observation details vary from instrument to instrument; the Keck Planet Finder \citep[KPF;][]{Rubenzahl2023:SoCal} and NEID \citep{Lin2022} utilize fast exposures (12~s and 55~s), while EXPRES \citep{LOST} exposes to a fixed signal-to-noise (S/N) and thus has variable exposure times (typically 140--178~s), and HARPS-N \citep{Phillips2016} takes 300~s exposures to average-out the 5.5~minute \textit{p}-mode oscillations. Exposure-aware models are therefore critical for unbiased inference from such data. These facilities obtain hundreds of spectra every day; jointly modeling all relevant timescales would involve a data volume on the order of $10^5$ observations per year. 

Since the naive (dense matrix) GP solution scales computationally as \bigO{N^3}, scalable methods are essential to fully utilize these large datasets, especially if we wish to extend such analyses to the spectral level ($\sim$$10^5$ pixels per observation). Consequently, much work has been devoted to scalable GP methods by recasting compatible kernels in terms of low-rank generators. For example, 
\celerite\ \citep{celerite} models utilize a basis set of complex exponentials which produce covariance matrices with semiseperable properties conducive to \bigO{N} linear algebra operations \citep{Ambikasaran2015}. Most commonly used GP kernels in astrophysics (e.g. half-integer Mat{\'e}rn family, or a stochastically-driven damped harmonic oscillator; SHO) can be represented with \celerite\ terms, and as such it has seen widespread adoption. \celerite\ models have even been extended to parallel (e.g., multiband) time series \citep{Gordon2020} as well as close-to-diagonal \citep[\project{S+LEAF};][]{SPLEAF,SPLEAF2} noise models. 

The general class of scalable GPs is encapsulated by state-space models (SSMs), which are defined for any GP kernel whose power spectral density (PSD) is a rational function \citep{HartikainenSarkka2010, Sarkka2013, SarkkaSDEbook}. For kernels with PSDs which are the sum of Lorentzians, \citet{Kelly2014} developed the CARMA model, which similarly achieves \bigO{N} solutions via a Kalman filter. Quasi-separable kernels also represent a structured subclass of SSMs \citep{Eidelman1999}; this connection powers the efficient \bigO{N} implementation in \tinygp\ \citep{tinygp} by using the SSM definition as a generator for quasi-separable matrices (QSMs). More directly though, the SSM definition can be itself be used to solve the GP regression problem in \bigO{N}. As a result, there are a number of packages which implement SSM/GP regression such as \project{BayesNewton}\footnote{\href{https://github.com/AaltoML/BayesNewton}{https://github.com/AaltoML/BayesNewton}} \citep{bayesnewton} and \project{Dynamax}\footnote{\href{https://github.com/probml/dynamax}{https://github.com/probml/dynamax}} in Python/JAX, and \project{TemporalGPs.jl}\footnote{\href{https://github.com/JuliaGaussianProcesses/TemporalGPs.jl}{https://github.com/JuliaGaussianProcesses/TemporalGPs.jl}} in Julia. While widely used in other fields outside astronomy, SSMs have seen some use in time domain astronomy for modeling stochastic signals in large survey time series \citep{Kelly2014}, jointly modeling multi-band light curves \citep{HuTak2020}, and detecting gravitational wave signals from pulsar timing \citep{Argus}.

Unfortunately, integrated measurements complicate these approaches, and no existing implementations handle this case. For the specific case of nonoverlapping exposures and constant exposure times, an exposure-integrated \celerite\ kernel function is also a \celerite\ kernel
(see Appendix~\ref{sec:celerite_integration} and the \texttt{TermConvolution} term in \celerite).
\citet{Miller2022} explored a way to scale approximate integrated kernels to large datasets using inducing points. However, none of these approaches can generate an exact covariance matrix for the general case of variable-length and/or overlapping exposures, due to the fact that the integrated kernel will either depend on the measurement-specific exposure length or will require careful logic to account for overlap. \citet{LuhnIntGP} developed the methodology for this general case by fully constructing the covariance matrix and solving with dense linear algebra. Though, for our motivating use case of EPRV solar datasets ($N \sim 10^5$), even computing a predictive mean and variance for a multicomponent kernel (i.e. a sum of kernels for each stochastic process of interest) is prohibitively expensive, let alone optimizing the hyperparameters. Thus, we seek a scalable solution without sacrificing model flexibility.

With this motivation, we turn to SSMs \citep[see][for a detailed introduction and equivalence to GPs]{SarkkaSDEbook, SarkkaBFSbook} which are equivalent descriptors of stochastic processes as GPs when their defining stochastic differential equation (SDE) is linear and driving noise is Gaussian. SSMs include all QSM kernels and can be generalized to arbitrary kernels at any desired precision by combining basis kernels \citep{Loper2020}, analogous to \celerite. As an example, \citet{SolinSarkka2014} show how to approximate the quasiperiodic kernel by constructing an SSM from its Taylor series expansion.

In this work, we show how an SSM augmented with an integral state yields an equivalent solution to the full integrated GP approach of \citet{LuhnIntGP} but is solved in linear time complexity and is compatible with parallel methods \citep[e.g.][]{Sarkka2021, Yaghoobi2025,Yaghoobi2025sqrt}. We introduce SSMs and their equivalence to GPs in Section~\ref{sec:SSM_GP_intro}. We then develop an augmented SSM with an integral state to accommodate integrated (and possibly overlapping) observations in Section~\ref{sec:integrated_ssm}. We implemented the full SSM-GP framework described in this paper in Python/\texttt{JAX} \citep{jax}, which we make available in the package State-space Models for O(Linear/log) GPs \citep[\smolgp\footnote{\href{https://github.com/smolgp-dev/smolgp}{https://github.com/smolgp-dev/smolgp}};][]{smolgp}. We demonstrate the equivalence and performance boost of this approach compared to traditional GP methods in Section~\ref{sec:benchmark}.
Finally, we summarize the method in Section~\ref{sec:conclusion} and conclude with other potential applications and future elaborations of this SSM-GP framework for more complex data or modeling requirements.

\section{GP\MakeLowercase{s} as State-space Models}\label{sec:SSM_GP_intro}

Let us orient ourselves with the definition of SSMs and their relationship to GPs. First, we will refresh ourselves with the GP problem statement (Section~\ref{sec:GP}) before introducing SSMs and their solution (Section~\ref{sec:SSM}), making the connections between these two complimentary frameworks explicit at each stage. We provide a summary of the components of an SSM and their analogy in the GP framework in Table~\ref{tab:ssm-gp-matrices}.

\begin{table*}
\centering
\caption{Summary of state-space model matrices and their relation to GPs}\label{tab:ssm-gp-matrices}
\begin{tabular}{|c|c|l|l|l|}
\hline\hline
Matrix & Size &  Name & Description & Analogy to GP \\
\hline
$\F$  &  $ d\times d$  &  Feedback matrix &  \parbox[t]{\tabcellwidth}{Governs the instantaneous deterministic dynamics of the latent state. The eigenvalues encode how perturbations to that state evolve (e.g. decay).}    &   \parbox[t]{\tabcellwidth}{Encodes the characteristic timescales of the process and the global correlation structure (decay, oscillations, periodicity, etc.).}  \\
\hline
$\L$ & $ d\times 1$ & Noise effect matrix & \parbox[t]{\tabcellwidth}{Maps the driving white noise to the latent state vector and its derivatives. It is simply $(0,...,1)$ for the cases we consider here (i.e. only the highest derivative term is driven by noise).} & \parbox[t]{\tabcellwidth}{How smooth the GP kernel is. If $\L$ has multiple nonzero entries, it is like a sum of GPs with varying smoothness.} \\
\hline
$\Qc$ & $ 1\times 1$ & Spectral density & \parbox[t]{\tabcellwidth}{Defines the amplitude of the white noise driving the process.} & \parbox[t]{\tabcellwidth}{Represents the power of the GP kernel's driving noise (related to the kernel's PSD amplitude).} \\
\hline
$\Pinf$ &  $ d\times d$ & Stationary covariance & \parbox[t]{\tabcellwidth}{Prior covariance of the latent state (i.e. before observing any data).} & \parbox[t]{\tabcellwidth}{Related to the GP kernel's (and its derivative(s)'s) amplitude.}  \\
\hline
\hline
$\H_k$ & $ D\times d$  & Observation model & \parbox[t]{\tabcellwidth}{Projects the latent $k^\text{th}$ state to the observed space.}       & \parbox[t]{\tabcellwidth}{Linear transformation from the latent process to the observed $f(t)$.}  \\
\hline
$\R_k$ & $ D\times D$ & Observation noise & \parbox[t]{\tabcellwidth}{The variance (and covariances, if $D>1$) of the measurement at time $t_k$.} & \parbox[t]{\tabcellwidth}{The $k^{th}$ element along the diagonal of the noise matrix $\R$.} \\
\hline
$\A_k$  & $ d \times d$ & Transition matrix & \parbox[t]{\tabcellwidth}{Maps the state vector forward by the time step $\Delta_k$.} & \parbox[t]{\tabcellwidth}{How the latent GP evolves deterministically over a time-lag $\Delta_k$.} \\
\hline
$\Q_k$  & $ d\times d$ & Process noise & \parbox[t]{\tabcellwidth}{Accumulated white noise injected into the state over one time step $\Delta_k$.}   & 	\parbox[t]{\tabcellwidth}{How the latent GP evolves stochastically over a time-lag $\Delta_k$.} \\
\hline
\end{tabular}
\\
{\; Note: Recall $d$ is the state ($\bm{x}$) dimension and $D$ is the data ($\bm{y}_n$) dimension; in this work we consider 1D time series, so $D=1$. $k$ is the state index which runs from 1 to $N$ for $N$ instantaneous data points; for $N$ integrated measurements, $k \in [1,2N]$.}
\end{table*}

\subsection{The GP problem statement}\label{sec:GP}

A temporal GP defines a probability distribution over functions of time $t$. For a GP with zero mean and covariance defined by the kernel function $k(t, t')$, these functions $f(t)$ are \citep{gpml}

\begin{align}
f(t) \sim \mathcal{GP}(0,\, k(t,t')).
\end{align}

Our measurements $\bm{y} = \{y_n\}_{n=1:N}$ taken at times $\bm{t} = \{t_n\}_{n=1:N}$ are noisy samples of the process,
\begin{align}
y_n = f(t_n) + \epsilon_n, 
\end{align}
where $\epsilon_n \sim \Normal(0,\sigma_n^2)$ is our Gaussian measurement noise with variance $\sigma_n^2$. Conditioning the GP on the observed data $(\bm{t},\bm{y})$ yields the predictive distribution
\begin{align}
f(t_\ast | \bm{t}, \bm{y}) \sim \mathcal{N}(\bm{\mu}_{GP}(t_\ast),\, \bm{\Sigma}_{GP}(t_\ast)).
\end{align}
with mean ($\mu$) and covariance ($\Sigma$) given by
\begin{equation}
\begin{aligned}
\label{eq:full GP solution}
\bm{\mu}_{GP}(\bm{t}_\ast) &= \bm{K}_\ast^T (\bm{K} + \bm{R})^{-1} \bm{y},  \\
\bm{\Sigma}_{GP}(\bm{t}_\ast) &= \bm{K}_{\ast\ast} - \bm{K}_\ast^T (\bm{K} + \bm{R})^{-1} \bm{K}_\ast.
\end{aligned}
\end{equation}

Above, $\bm{K}$ is the covariance matrix computed from the kernel function for all pairs of observations, i.e. $K_{ij} = k(t_i,t_j)$ for $i,j \in [1,N]$. $\bm{K}_\ast$ denotes the kernel function evaluated between the data and an arbitrary set of $M$ ``test'' points, i.e. $\bm{K}_{ij} = k(t_{i}, t_{\ast j})$ for $i \in [1,N],\,j \in [1,M]$. Likewise, $\bm{K}_{\ast\ast}$ is the same for all pairs of test points. $\bm{R}$ is the matrix of measurement covariances; in the usual case of independent but heteroskedastic errors, $\bm{R} = \text{diag}(\{\sigma_n^2\}_{n=1:N})$.

The kernel function, which itself usually depends on some number of ``hyperparameters'' ($\bm{\theta}$), can have many different functional forms as long as $\bm{K}$ is positive definite. We often work with stationary kernels, meaning they are functions only of the time difference $\Delta \equiv |t - t'|$. That is, we have $k(\Delta | \bm{\theta})$ instead of $k(t,t'|\bm{\theta})$. 

Eq.~\ref{eq:full GP solution} suffers from the curse of dimensionality, scaling computationally as \bigO{N^3} for naive $N\times N$ matrix inversion and multiplication. Even holding these matrices in memory scales as \bigO{N^2}, which can be $100+$~GB for $\gtrsim10^5$ data points. While QSM methods can reduce both to \bigO{N} for compatible kernels, we seek an approach compatible with the general case of integrated (with variable exposure time) and sometimes overlapping measurements

\subsection{The SSM problem statement}\label{sec:SSM}

See \citet{SolinSarkka2014} for an introduction to reformulating temporal GP regression into a linear Gaussian SSM, which we summarize here. We also provide a full worked example in Appendix~\ref{appendix:sho} for an SHO. We note here that we define the subscript $n$ as the index for the data points, i.e. $\bm{y} = \{y_n\}_{n=1:N}$, and the subscript $k$ for the discretized states, i.e. $\bm{x} = \{x_k\}_{k=1:K}$. In this section, $n$ and $k$ can be used interchangeably as we discretize the model at the data points. However, in Section~\ref{sec:integrated_ssm} where we consider integrated measurements, we will discretize the model at the exposure start and end times, and so in that case we will have $K = 2N$ states.

In the state-space formalism, we can think of these functions $f(t)$ as instead solving a $d$-th order linear SDE driven by a white noise process $w(t)$,

\begin{equation}
\begin{aligned}
\label{eq:sde}
\frac{\dd\bm{x}(t)}{\dd t} &= \F \bm{x}(t) + \L \bm{w}(t), \\
f(t) &= \bm{H} \bm{x}(t),
\end{aligned}
\end{equation}
where we now have a length-$d$ vector $\bm{x}(t) = (x, \dot{x}, \ddot{x}, ...)^T$ of the latent process $x(t)$ and its first $d-1$ time derivatives. $\F$, called the feedback matrix, is a $d\times d$ matrix encoding the state evolution. The noise effect matrix, $\L$, is a column vector that determines which of the state's derivatives are driven by the white noise process $\bm{w}(t)$, which is itself defined by its spectral density $\Qc$,
\begin{align}\label{eq:Qc_from_w(t)}
\text{E}[\bm{w}(t)\bm{w}(t')^T] = \Qc\delta_\text{dirac}(t-t').
\end{align}

Finally, the observation matrix $\H_k$ projects the latent state vector at given time $\bm{x}(t_k)$ to the observed space $f(t_k)$. When dealing with instantaneous GP kernels, and if only the state $x$ is measured, then $\H_k$ is simply a constant row vector $[1, 0,...,0]$ that picks out the latent state $x$. Provided they exist, derivatives of $x(t)$ can be observed by simply changing the corresponding elements of $\H_k$ to a 1. Likewise, if we are interested in the sum of the latent state and its first derivative, the observation matrix becomes $\H_k = [1, 1, 0,\dots, 0]$. Examples of such a GP model are explored in \citealt{Aigrain2012}, \citealt{Rajpaul2015}, \citealt{Jones2017}, \citealt{Gilbertson2020b}, and \citealt{Tran2023}. In general, $\H_k$ may vary from data point to data point, or might include an amplitude term (which could be different depending on which instrument made the measurement). We can also apply linear operators to $\H$ and preserve the linear Gaussian behavior of the SSM \citep{Sarkka2013}.

Like before, our measurements\footnote{Here we assume 1D data, i.e. $y_n$ is a scalar. Though, as long as $\H$ etc. (see Table~\ref{tab:ssm-gp-matrices}) are appropriately shaped, multivariate data with dimension $D$ is perfectly compatible here.} are noisy samples of the observed latent process
\begin{align}\label{eq:measurement_model}
y_n = f(t_n) + \epsilon_n = \bm{H_n} \bm{x}(t_n) + \epsilon_n, 
\end{align}
where the measurement noise $\epsilon_n$ is the same as before.

\subsubsection{The GP--SSM equivalence}\label{sec:GP-SSM_equiv}

With the observation model in the state-space defined, the question now is how to go from a GP kernel function to an SDE (Eq.~\ref{eq:sde}). Here, we provide a top-level overview; for more details and an example derivation for the SHO kernel, see Appendix~\ref{appendix:sho}. Section~4 in \citet{HartikainenSarkka2010}, which does the same for the Mat{\'e}rn family and squared exponential kernels, is also useful.

For a desired GP kernel, the matrices $\F$, $\L$, and $\Qc$ are derived from the PSD, $S(\omega)$, which is the Fourier dual of the kernel function \citep[per the Wiener–Khinchin theorem;][]{gpml}
\begin{equation}\label{eq:psd}
    S(\omega)=\int_{-\infty}^{\infty}k(\Delta)e^{-i\omega\Delta}\dd\Delta.
\end{equation}

If the PSD is a rational function, then we can write it as the output of a linear time-invariant (LTI) system where the input is white noise with amplitude $q$ \citep[Theorem 3.4 in][]{optimalfiltering},
\begin{align}\label{eq:transfer}
    S(\omega) = |H(i\omega)|^2 q,
\end{align}
where $H(i \omega)$, called the transfer function, determines how the system responds to the white noise input. The appropriate transfer function $H(s)$, in the complex $s$-domain, can be identified through spectral factorization \citep[see Ch. 9 in][]{optimalfiltering}. By definition, the transfer function is the ratio of the output to the input of a system in the frequency domain,
\begin{align}\label{eq:ratio}
    H(s) \equiv \frac{X(s)}{W(s)},
\end{align}
with $X(s)$ and $W(s)$ being the Laplace transforms of $x(t)$ and $w(t)$ (the output and the input, respectively). One can choose $q$ such that $H(s)$ can be written as 
\begin{align}\label{eq:H_as_denominator}
    H(s) = \frac{1}{s^d + a_{d-1}s^{d-1} + \dots + a_0},
\end{align}
where $a_i$ are coefficients.
This form is convenient because it lets us rearrange Eq.~\ref{eq:ratio} into
\begin{align}\label{eq:laplace_sde}
    W(s) = s^d F(s) + a_{d-1} s^{d-1} F(s) + \dots + a_0 F(s),
\end{align}
which can be converted to the time domain by inverse Laplace transforming both sides, turning multiplication by $s$ in the Laplace domain into time derivatives in the time domain, thus giving the SDE
\begin{align}\label{eq:linear_sde}
    \frac{\dd^d x(t)}{\dd t^d} + a_{d-1} \frac{\dd^{d-1} x(t)}{\dd t^{d-1}} + \dots a_0 x(t) = w(t).
\end{align}
Finally, the SDE can be put into matrix form by constructing the matrices $\F$ and $\L$:
\begin{align}\label{eq:generic_F}
    \F = \begin{pmatrix}    
        0 & 1 & 0 & \dots & 0\\
        0 & 0 & 1 & \dots & 0\\
   \vdots & \vdots & \ddots & \ddots & \vdots \\
    -a_0 & -a_1 & \dots & \dots & -a_{d-1} 
            \end{pmatrix}, \; 
    \L = \begin{pmatrix}
                0 \\ 0 \\ \vdots \\ 1
            \end{pmatrix},
\end{align}
yielding Eq.~\ref{eq:sde}.

\subsubsection{Solution to the SDE}

To solve the SDE, we first need to define the initial conditions of the process. These are determined by the mean of the process (which we take to be zero without loss of generality) and the prior covariance, also called the stationary covariance $\Pinf$. $\Pinf$ is defined as the covariance after the system has settled, i.e. $t \rightarrow \infty$ so $d\P/dt=0$. It has the following definition in the state-space formalism as the solution to the continuous-time Lyapunov equation \citep[see Chapters 5.5, 6.1, \& 6.5 in][for the derivation from the expected covariance of the SDE in the stationary limit]{SarkkaSDEbook},
\begin{align}\label{eq:lyapunov_eq}
\frac{\dd\bm{P}}{\dd t} = \F \Pinf + \Pinf \F^T + \L \Qc \L^T = 0. 
\end{align}

Given $\F$ and $\L$, one can solve Eq.~\ref{eq:lyapunov_eq} for the elements of $\Pinf$ in terms of $\Qc$. The result should be a diagonal matrix. The first element corresponds to the stationary covariance of the latent process, i.e. the kernel function evaluated at zero time-lag, $k(0)$. This equivalence allows one to define $\Qc$ in terms of the kernel parameters, and subsequently $\Pinf$ is determined. One will also find that the lone element of $\Q_c$ is the same $q$ from Eq.~\ref{eq:transfer}. See Appendix~\ref{appendix:sho} for a worked example with the SHO kernel.

The solution to Eq.~\ref{eq:sde} at time $t$ relative to another time $t_0$ is
\begin{align}\label{eq:sde_sol}
    \bm{x}(t) = \bm{\Phi}(t-t_0)\bm{x}(t_0) + \int_{t_0}^t \bm{\Phi}(t-\tau)\L w(\tau) \dd\tau,
\end{align}
where the transition matrix 
\begin{align}\label{eq:phi}
\bm{\Phi}(t-t_0) = \exp(\F (t-t_0))    
\end{align}
is the matrix exponential of the feedback matrix times the relative time difference. To determine the conditional solutions given the data, we can discretize Eq.~\ref{eq:sde_sol} at time $t_{k+1}$ relative to $t_k$ as
\begin{align}\label{eq:sde_sol_discrete}
\bm{x}_{k+1} = \A_k \bm{x}_k + \bm{q}_k, \quad \bm{q}_k \sim \Normal(\bm{0},\Q_k) 
\end{align}
where $\bm{x}_k \equiv \bm{x}(t_k)$, $\A_k = \bm{\Phi}(\Delta_k)$, and $\Delta_k = t_{k+1}-t_k$. The matrix $\Q_k = E[\bm{q}_k \bm{q}_k']$ is called the process noise covariance and is given by substituting Eq.~\ref{eq:Qc_from_w(t)} into the second term of Eq.~\ref{eq:sde_sol}
\begin{align}\label{eq:Q}
\Q_k &= \int_0^{\Delta_k} \bm{\Phi}(\Delta_k - \tau) \L \Qc \L^T \bm{\Phi}(\Delta_k - \tau)^T \dd\tau \nonumber \\
    &= \Pinf - \A_k \Pinf \A_k^T.
\end{align}

The last equality, which can be found in \citet{SolinSarkka2014b} and \citet{SarkkaSDEbook}, is preferred over working out the Lyapunov integral for its speed and numerical stability, especially when $\Pinf$ and $\A_k$ are known analytically. Alternatively, the integral can be efficiently computed by taking the matrix exponential of a block upper-triangular matrix involving $\F$, $\L \Qc\L^T$, and $\Delta$, from which the product of two submatrices gives exactly $\Q_k$ \citep[][see Eq.~\ref{eq:Q_from_vanloan} for an example]{VanLoan1978}. However, for $\Delta$ around $\sim$$10^3$ times longer than the kernel timescale, the numerical value of this exponential is too large to be represented numerically; for data with large gaps, the analytic forms are required.

For a pair of points $t$ and $t'$, the GP covariance function is \citep[Eq. 4 in][]{HartikainenSarkka2010}
\begin{align}\label{eq:ssm-kernel-function}
k(t,t') =
\begin{cases}
    \H(t') \Pinf \bm{\Phi}(|t-t'|)^T \H(t)^T, & t \geq t', \\
    \H(t) \bm{\Phi}(|t-t'|) \Pinf \H(t'), & t < t'.
\end{cases}
\end{align}

To generate predictions or compute likelihoods, because everything is Gaussian, we need only track the means and variances of these $\bm{x}_k$ from state to state. A vast literature anchored in control theory exists for doing just that as countless real-world problems can be modeled as state-space systems. The Kalman filter and Rauch--Tung--Striebel (RTS) smoothing algorithms, discussed further in the following sections, yield the optimal predictions for such linear Gaussian systems; these optimal predictions are identical to the GP conditioned mean and variance. See \citet{SarkkaBFSbook} \citep[][for the continuous case]{SarkkaSDEbook} for the full derivations or Figure~\ref{fig:ss_vs_gp} for a numerical demonstration.

\subsection{GP conditioning}\label{sec:kalman/rts}

The problem of conditioning the GP is to derive the optimal prediction for the mean, $\bm{m}$, and covariance, $\P$, of the state $\bm{x}$ given the data $\bm{y}$. For a linear Gaussian SSM, the solution is given by Bayesian filtering and smoothing algorithms \citep{SarkkaBFSbook}. Formally, for a 1D time series, these algorithms scale as \bigO{N d^3} because they involve matrix operations (multiplications, inversions) on (at largest) $d \times d$ matrices per iteration through the $N$ data points. For the SSMs we consider, $d = 2$ or 3, so for realistic datasets $N \gg d$ and so \bigO{N d^3}~$\simeq$~\bigO{N}.

\subsubsection{Kalman filter}\label{sec:kalman}

The Kalman filter \citep{Kalman1960} computes the conditional probabilities at each state given all previous data, $p(\bm{x_k}|\bm{y}_{1:k})$. Since everything is linear and Gaussian, we just need to compute the mean and covariance at each state, iterating chronologically through the data. To start the iteration, we initialize the state to $\bm{x}_0 \sim \Normal(\bm{m}_0,\bm{P}_0)$ with $\bm{m}_0$ set to the kernel mean function (usually zero for the state and all its derivatives) and $\bm{P}_0$ set to the stationary covariance ($\Pinf$). The algorithm is then \citep[Theorem 6.6 in][]{SarkkaBFSbook}

\begin{align}
\text{Prediction:}&  \nonumber \\
\m_{k}^{-} &= \A_{k-1} \m_{k-1}, \nonumber \\
\P_{k}^{-} &= \A_{k-1} \P_{k-1} \A_{k-1}^T + \Q_{k-1}. \label{eq:kalman_predict} \\
\text{Update:}&   \nonumber \\
v_k &= y_k - \H_k \m_k^{-}, \nonumber \\
\S_k &= \H_k \P_k^{-} \H_k^T + \R_k, \nonumber \\
\K_k &= \P_k^{-} \H_k^T \S_k^{-1}, \nonumber \\
\m_k &= \m_k^{-} + \K_k v_k, \nonumber \\
\P_k &= \P_k^{-} - \K_k \S_k \K_k^T. \label{eq:kalman_update}
\end{align}

The prediction step (Eq.~\ref{eq:kalman_predict}) can be thought of as simply transitioning from the previous (filtered) state to the current state. The update step (Eq.~\ref{eq:kalman_update}) first calculates the ``surprise term'' $v_k$ (also called the innovation), which is simply the difference between our prediction in the observed space and the actual measured value. The uncertainty in the prediction, $\S_k$ (also called the innovation covariance), defines the Kalman gain ($\K_k$) which effectively weights the surprise term for the purpose of updating our predicted mean $\m_k$ and variance $\P_k$. The filtered $\m_k$ and $\P_k$ are then carried to the next iteration, continuing until $k=N$. Note that these matrices are all in the state-space. That is, the $\m$s have shape $d \times 1$ and the $\P$s have shape $d \times d$. Only the innovation step projects the mean state to the observed space ($\H_k \m_k$) to compare to the measured $y_k$.

\subsubsection{RTS smoother}\label{sec:rts}

The RTS \citep{RTS} smoothing algorithm refines these predictions using future data to give $p(\bm{x}_k|\bm{y}_{1:N})$ for all $k$ by applying the transition matrix to the Kalman filter results in reverse-chronological order. It is important to note that this is not the same as Kalman filtering in reverse as causality assumptions in the noise model must be obeyed (e.g. driving and damping backwards in time is not the same as undriving and undamping).

The final Kalman-filtered state is already informed by all other data points; as such, it is already smoothed. That is, $\hat{\m}_N = \m_N$ and $\hat{\P}_N = \P_N$. Then, starting from the penultimate ($k=N-1$) state  and iterating in reverse-chronological order, the RTS algorithm is \citep[Theorem 12.2 in][]{SarkkaBFSbook}
\begin{align}
\m_{k+1}^{-} &= \A_k \m_k, \nonumber \\
\P_{k+1}^{-} &= \A_k \P_k \A_k^T + \Q_k, \nonumber \\
\G_k &= \P_k \A_k^T \left[ \P_{k+1}^{-} \right]^{-1}, \nonumber \\
\hat{\m}_k &= \m_k + \G_k \left[ \hat{\m}_{k+1} - \m_{k+1}^{-} \right], \nonumber \\
\hat{\P}_k &= \P_k + \G_k \left[ \hat{\P}_{k+1} - \P_{k+1}^{-} \right] \G_k^T.  \label{eq:rts}
\end{align}

Note $\m_k$, $\P_k$, $\m_{k+1}^{-}$, and $\P_{k+1}^{-}$  are already computed by the Kalman filter (Eqs.~\ref{eq:kalman_predict}, \ref{eq:kalman_update}). $\G_k$ is the ``smoothing gain'' which weights ``how much we got the $k+1^{th}$ state's prediction wrong'' by the data-conditioned variance in the process between $k+1 \rightarrow k$, to ``correct'' the Kalman-filtered state $k$. The smoothed results are mathematically equivalent to full GP conditioning (Eq.~\ref{eq:full GP solution}) when the state-space SDE is linear and the driving noise is Gaussian \citep[see Ch. 12.4 of][and references therein]{SarkkaSDEbook}. That is, $\H \hat{\m} \equiv \bm{\mu}_{GP}$ and $\H \hat{\P} \H^T \equiv \bm{\Sigma}_{GP}$.


\subsection{The log-likelihood}

Byproducts of the Kalman filter are the ingredients to compute the likelihood \citep[Eq. 16.5 in][]{SarkkaBFSbook},
\begin{align}
p(\bm{y}_{1:N} | \bm{\theta}) = \prod_{n=1}^{N} p(y_n|\bm{y}_{1:n-1}, \bm{\theta}).
\end{align}
The individual $p(y_n|\bm{y}_{1:n-1},\bm{\theta})$ are given by the first two pieces of the ``update'' step in Eq.~\ref{eq:kalman_update}, since
\begin{align}
p(y_n|\bm{y}_{1:n-1},\bm{\theta}) = N(\H_n \m_n^{-}, \S_n).
\end{align}
The total log-likelihood $\mathcal{L}(\bm{y}|\bm{\theta}) = \log p(\bm{y}_{1:N}|\bm{\theta})$ is thus
\begin{align}\label{eq:loglikelihood}
\mathcal{L}(\bm{y}|\bm{\theta}) = -\frac{1}{2}\sum_{n=1}^N \left( \log\det(2\pi \S_n) + \v_n^T \S_n^{-1} \v_n \right).
\end{align}

Eq.~\ref{eq:loglikelihood} can be input to the usual methods (e.g. gradient descent, Markov chain Monte Carlo) for hyperparameter optimization.

\subsection{Predicting at arbitrary times}\label{sec:arbitrary_predicting}

It is straightforward to extend the Kalman/RTS algorithms to predict the mean and covariance at an arbitrary time $t_\ast$. This is related to the idea of ``fast sampling'' described in Section 4.6 of \citet{SarkkaBFSbook}. There are three cases:

\begin{enumerate}

\item \textit{Retrodiction}. ($t_\ast < t_1$): If the test point is before the data, compute the RTS smoothed estimate at $t_\ast$ from the first measurement's smoothed ($\hat{\m}_1,\,\hat{\P}_1$) and predicted ($\m_1^-,\,\P_1^-$) states. The ``filtered'' mean at the test point, $\m_\ast$, is set to the kernel mean (e.g. zero) and the covariance $\P_\ast$ is the stationary covariance $\Pinf$.

\item \textit{Interpolation}. ($t_1 < t_\ast < t_N$): If the test point is during the data, we use the usual Kalman prediction step from the most recent data point's Kalman-filtered state and then refine the prediction with a RTS smoothing step from the nearest future data point's smoothed state.

\item \textit{Extrapolation}. ($t_\ast > t_N$): If the test point is after the data, we simply use the Kalman prediction from the final data point's filtered/smoothed state.

\end{enumerate}

\begin{algorithm}
\caption{SSM prediction at arbitrary time $t_\ast$}\label{alg:predict}
\begin{algorithmic}[1]
\State $t_1 \gets \min(t)$
\State $t_N \gets \max(t)$
\If{$t_\ast < t_1$} \Comment{Retrodict}
    \State $\Delta_\ast \gets t_1 - t_\ast$
    \State $\A_\ast \gets \Phi(\Delta_\ast)$
    \State $\m_0 \gets \bm{0}$
    \State $\P_0 \gets \Pinf$
    \State $\G_0 \gets \P_0 \A_\ast^T [\P_1^-]^{-1}$
    \State $\hat{\m}_\ast \gets \m_0 + \G_0[\hat{\m}_1 - \m_1^-]$
    \State $\hat{\P}_\ast \gets \P_0 + \G_0[\hat{\P}_1 - \P_1^-]$
\ElsIf{$t_1 < t_\ast < t_N$} \Comment{Interpolate}
    \State $t_\text{prev} \gets \max(t | t<t_\ast)$
    \State $t_\text{next} \gets \min(t | t>t_\ast)$
    \State $\A_\text{prev} \gets \bm{\Phi}(t_\ast - t_\text{prev})$
    \State $\A_\text{next} \gets \bm{\Phi}(t_\text{next} - t_\ast)$
    \State $\Q_\text{prev} \gets \Q(t_\ast - t_\text{prev})$
    \State $\m_\ast \gets \A_\text{prev} \m_{k-1}$
    \State $\P_\ast \gets \A_\text{prev} \P_{k-1} \A_\text{prev}^T + \Q_\text{prev}$
    \State $\G_\ast \gets \P_0 \A_\text{next}^T [\P_1^-]^{-1}$
    \State $\hat{\m}_\ast \gets \m_\ast + \G_\ast[\hat{\m}_\text{next} - \m_\text{next}^-]$
    \State $\hat{\P}_\ast \gets \P_\ast + \G_\ast[\hat{\P}_\text{next} - \P_\text{next}^-]$
\ElsIf{$t_\ast > t_N$} \Comment{Extrapolate}
    \State $\Delta_\ast \gets t_\ast - t_N$
    \State $\A_\ast \gets \bm{\Phi}(\Delta_\ast)$
    \State $\Q_\ast \gets \Q(\Delta_\ast)$
    \State $\hat{\m}_\ast \gets \A_\ast \m_N $
    \State $\hat{\P}_\ast \gets \A_\ast \P_N \A_\ast^T + \Q_\ast$
\EndIf
\end{algorithmic}
\end{algorithm}

Algorithm~\ref{alg:predict} demonstrates this in pseudocode. An alternative algorithm was described in the appendix of \citet{Kelly2014} using a linearized form for the predicted state at the test point to derive their smoothing equations. Though it is significantly more computationally expensive because it recomputes the smoothing gain at every test point, which in each case involves a loop through all the future data to that test point. 

In the GP framework, predicting at $M$ arbitrary test points involves rectangular matrix multiplications; these can still be \bigO{N} for QSMs (e.g. Algorithm 5 in \citealt{Pernet2017}, see also \citealt{Pernet2023}). Generally, Eq.~\ref{eq:full GP solution} for GPs (QSM or dense) scales linearly with the number of test points $M$, though the implicit conditioning still grows with $N^2$ (or $N$ for QSMs). By contrast, Algorithm~\ref{alg:predict} is embarrassingly parallel; each prediction only depends on the nearest (past and/or future) data point. It is also significantly less memory intensive (see e.g., Figure~\ref{fig:benchmark}).

\begin{figure*}
    \centering
    \includegraphics[width=\linewidth]{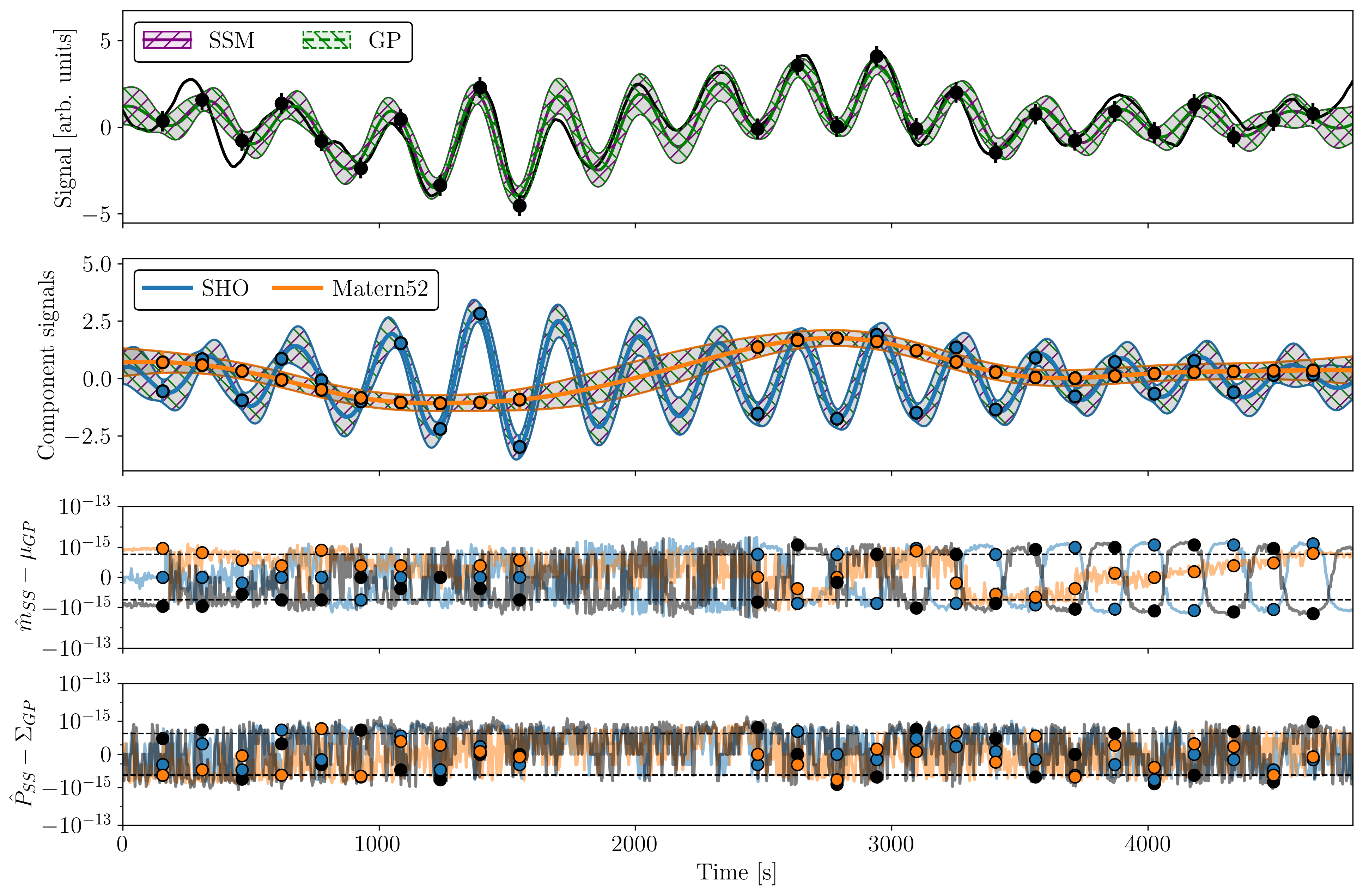}
    \caption{Equivalence of linear Gaussian SSMs to GPs. \textbf{Top}: The true signal (in black) is the sum of an SHO and Mat{\'e}rn-5/2 kernel. Synthetic measurements (black points) are noisy samples of this process. The purple ``/'' band shows the predicted mean and 1$\sigma$ variance given the data from the SSM method; the green ``\textbackslash'' band shows the same for the full GP method. \textbf{Middle}: The same as the top panel but decomposed into each component kernel's prediction at the test points (filled bands) and at the data (points). \textbf{Bottom}: The bottom two panels show the residuals for the predicted mean and variance from the overall model prediction (black line) and component predictions (blue/orange lines) at the test points, and well as at the data (points), between the SSM method and the full GP method. The horizontal dashed lines denote machine precision for 64-bit floating-point numbers in the dynamic range of the data; In all cases, the difference between \smolgp\ and \tinygp\ is within this level.}
    \label{fig:ss_vs_gp}
\end{figure*}

To demonstrate the SSM--GP equivalence, we numerically validated the SSM approach \citep[using \smolgp;][]{smolgp} to the GP approach \citep[using \tinygp;][]{tinygp} for a multicomponent (see Appendix~\ref{appendix:multicomponent}) kernel made from the sum of an SHO and a Mat{\'e}rn-5/2. We took a random realization of this process as our true signal, from which we generated synthetic measurements. We then conditioned an SSM and GP on these data and made predictions at a high-resolution test grid. The results of conditioning and predicting (for the full model and for each component kernel), as well as the log-likelihood, were equivalent to within machine precision. Figure~\ref{fig:ss_vs_gp} shows this comparison.

\section{Integrated Observations}\label{sec:integrated_ssm}

\subsection{In the GP framework}\label{sec:integrated_gp}

Handling exposure-integrated observations, especially when observations are allowed to overlap, requires significant bookkeeping. In the traditional GP framework, the cross-covariance between two observations at times $t_1$ and $t_2$ with corresponding exposure times $\delta_2$ and $\delta_2$ is
\begin{align}
k_{FF}(t_1,t_2,\delta_1,\delta_2) = \frac{1}{\delta_1 \delta_2} \int_{t_1-\delta_1/2}^{t_1+\delta_1/2} \int_{t_2-\delta_2/2}^{t_2+\delta_2/2} k(t,t') \dd t \dd t'. 
\end{align}
For the cases of fully separated or perfectly overlapping observations (such as an observation with itself, i.e. the diagonal), \citet{LuhnIntGP} (hereafter \citetalias{LuhnIntGP}) derived a closed-form analytic solution for the SHO kernel. These two cases can then be combined with appropriate weighting along the nonoverlapping and shared segments of partially overlapping observations to correctly recover the joint covariance for arbitrary overlap. However, the logic to implement this in practice prevents the use of quasi-separable linear algebra, even if the underlying kernel is stationary and quasi-separable, for the simple reason that while the overlapping observations live near to the diagonal (in a time-ordered matrix), they lack an exploitable structure; they also do not form a LEAF \citep{SPLEAF} matrix. 

A datastream from a single instrument, usually\footnote{An infrared echelle spectrograph might nondestructively (``up-the-ramp'') read-out during an exposure up to a fixed S/N independently across a number of pixel channels, yielding a different exposure time for each spectral segment. Additionally, adjacent echelle orders often cover overlapping wavelength ranges, so in such a scenario there can be simultaneous, overlapping measurements of the same signal from the same detector.}, does not have any overlapping observations. If so, and if the exposure time is constant for all data points, it can in some cases be possible to represent the integrated GP kernel in quasi-separable form. For example, an integrated SHO kernel can be expressed as the sum of four SHO kernels at shifted times and modified amplitudes, plus two constant terms, each of which is compatible with a quasi-separable form. However, combining data from multiple instruments during shared observing windows will often yield overlapping pairs of observations. In that case, the GP formalism is cursed to rely on dense matrix computations. In general though, for realistic datasets, overlapping observations represent a small fraction of the total number of data points (and certainly a small fraction of all the possible pairs of observations), so the dense matrix representation (while formally correct) is an inefficient way to deal with this problem.

\subsection{In the SSM framework}\label{sec:augmented_ssm}

SSMs are commonly used to model binned measurements in the context a fast-rate process that is occasionally sampled at a ``slow-rate,'' where each observed sample is the average of some number of (unobserved) fast-rate samples \citep[e.g.,][]{Guo2015, Fatehi2017, Salehi2018}, or as the integral of a continuous process during the interval between measurements \citep{Qian2021}. In our case, we wish to draw an exact equivalence to \citetalias{LuhnIntGP}, i.e. we have infrequent integrated samples of a continuous function where our measurements may overlap with one another. Our approach, based on footnote 1 of \citet{Yaghoobi2025}, essentially converts the method of \citetalias{LuhnIntGP} into an \bigO{N} algorithm by integrating the dynamics rather than the covariance.

Let $y_n$ be the exposure-averaged measurement at observation $n \in [1,N]$, 
\begin{align}
y_n &= \frac{1}{\delta_n} \int_{t^s_n}^{t^e_n} \H_n \bm{x}(t) \dd t + \epsilon_n.  
\end{align}
where $t^s_n$ and $t^e_n$ are the start and end times of the exposure, which has length $\delta_n = t^e_n - t^s_n$. Because integration is a linear operator, the measurement model remains linear and so is still described by a SSM \citep{Sarkka2013}. Moreover, our state is already the joint state of the instantaneous latent state and its first $d-1$ time derivatives. We can thus simply introduce the integral state $z$ such that
\begin{align}
\frac{\dd z}{\dd t} = x. 
\end{align}

We can then augment the SSM to be of the joint state $\xaug = [\bm{x};z]$, where $\bm{x}$ is as before. The augmented SDE becomes
\begin{align}\label{eq:augmented_sde}
\frac{\dd \xaug}{\dd t} &= \Faug \xaug(t) + \Laug \bm{w}(t),
\end{align}
with the augmented matrices
\begin{align}
\Faug = \begin{pmatrix} \F & \begin{matrix} 0 \\ 
     \vdots \end{matrix} \\
        \begin{matrix} 1 & 0\dots \end{matrix} &  \end{pmatrix}, \quad 
\Laug =  \begin{pmatrix}\L \\ 0 \end{pmatrix}.
\end{align}
The augmented state dimension becomes $d+1$. {\bfseries
One can verify this gives the correct system of SDEs,
\begin{align}
    \frac{\dd \xaug}{\dd t} = \begin{pmatrix}
        \frac{\dd\bm{x}}{dt} \\ \frac{\dd z}{\dd t}
    \end{pmatrix} = \Faug \xaug + \Laug \bm{w} = \begin{pmatrix}
        \F \bm{x} \\ x
    \end{pmatrix} + \begin{pmatrix}
        \L \bm{w} \\ 0
    \end{pmatrix}.
\end{align}
}
The corresponding augmented transition matrix is
\begin{align}\label{eq:phiaug}
\Phiaug(\Delta) = \begin{pmatrix}
                    \bm{\Phi}(\Delta) & 0 \\ 
                    \Phibar_x(\Delta) & 1 \end{pmatrix}, 
        \quad \Phibar(\Delta) = \int_0^\Delta \bm{\Phi}(t) \dd t,
\end{align}
where $\Phibar_x(\Delta)$ is a $1\times d$ vector containing the integral, over the interval $\Delta$, of the top row of the transition matrix; i.e., the integral of the transition in $x$ and each of its derivatives. In other words, $\bm{x}$ evolves according to the instantaneous dynamics like normal, while $z$ is accumulated over the transition. We can verify that Eq.~\ref{eq:phiaug} gives the correct behavior by checking the solution to the SDE matches Eq.~\ref{eq:sde_sol},
{\small
\begin{align}
    \xaug(t) &= \Phiaug(t-t_0)\xaug(t_0) + \int_{t_0}^t \Phiaug(t-\tau)\Laug w(\tau) \dd\tau \nonumber \\
   \begin{pmatrix}
       \bm{x}(t) \\ z(t)
   \end{pmatrix} &= \begin{pmatrix}
        \bm{\Phi}(t-t_0)\bm{x}(t_0) + \int_{t_0}^t \bm{\Phi}(t-\tau)\L w(\tau) \dd\tau \\ 
        z(t_0) + \Phibar(t-t_0)\bm{x}(t_0) + \int_{t_0}^t \Phibar(t-\tau)\L w(\tau) \dd\tau 
        \end{pmatrix}.
\end{align}
}
In practice, one can simply compute $\Phiaug(\Delta)$ numerically as the matrix exponential of $\Faug \Delta$. However, it is more numerically stable to use the method of \citet{VanLoan1978} to get $\Phibar(\Delta) = \bm{G}_3$, where
\begin{align}
\bm{C} = \begin{pmatrix}\F & \I \\ \Z & \Z\end{pmatrix} \; \rightarrow \; 
\exp(\bm{C} \Delta) = \begin{pmatrix}\bm{F}_3 & \bm{G}_3 \\ \Z & \bm{F}_4\end{pmatrix},
\end{align}
where $\bm{F}_3$ and $\bm{F}_4$ are other block matrices irrelevant to the definition of $\Phibar$. That is, one can extract only $\bm{G}_3 \equiv \Phibar(\Delta)$ and then assemble $\Phiaug$ via Eq.~\ref{eq:phiaug}. Of course, for certain kernels, the $\Phibar$ integral may be doable analytically (e.g., Eq.~\ref{eq:sho_phibar}); this is preferred whenever possible for numerical stability over long transitions.

The augmented process noise is 
\begin{align}\label{eq:Q_aug}
\Qaug &= \int_0^\Delta \Phiaug(t)\Laug \Qc\Laug^T\Phiaug(t)^T \dd t \\
        &= \begin{pmatrix}\int_0^\Delta \bm{\Phi} \L \Qc \L^T \bm{\Phi}^T \dd t & \int_0^\Delta \bm{\Phi} \L \Qc \L^T \Phibar^T \dd t \\
            \int_0^\Delta \Phibar \L \Qc \L^T \bm{\Phi}^T \dd t  &  \int_0^\Delta \Phibar \L \Qc \L^T \Phibar^T \dd t\end{pmatrix}, \nonumber
\end{align}
where $\bm{\Phi}$ and $\bm{\bar{\Phi}}$ are both functions of $t$, though we omitted writing that explicitly for notational compactness. Each of these components has its own representation as a Van Loan matrix exponential \citep{VanLoan1978}, or $\Qaug$ can be computed at once from $\Faug$ and $\Laug \Qc\Laug^T$ (see Eq.~\ref{eq:Q_from_vanloan}). In practice, because the matrix exponential is numerically unstable for large $\Delta$ (i.e., long gaps in the data set), we do the former, so that the top-left block ($\Q$ for the base SSM) can be populated by its analytic solution. While the other blocks are unstable for large $\Delta$, in practice they are immediately reset at the end of the gap so the result is unaffected. If there are indeed long ($\Delta \gtrsim 10^5$ time units) integration intervals in the measurements, then Eq.~\ref{eq:Q_aug} must be defined analytically by working out the integrals; the identity in Eq.~\ref{eq:Q} involving $\Pinf$ and $\A$ cannot be used to obtain $\Qaug$ because the $z$ state is a zero-variance process (i.e., $\Faug$ is noninvertible).

With the augmented model fully defined, we can discretize at the starts and ends of every exposure, yielding $K = 2N$ total states in total. Only the states $\xaug_k$ such that $t_k$ corresponds to the end of an exposure (with matching timestamp $t^e_n$) will have measurements ($y_n$) for the usual Kalman update step. Conversely, the states $\xaug_k$ such that $t_k$ is the start of an exposure (with timestamp $t^s_n$) will have its $z$ state reset to zero; this is effectively the ``update'' step in that we ``measure'' nothing at the start of an exposure. That is, the measurement model is
\begin{equation}
\begin{aligned}\label{eq:augmented_observation_model}
0 &= z(t_k) & \text{if} \; t_k = t^s_n \; \text{is an exposure start}, \\
y_n &= \Haug_n \xaug_k + \epsilon_n  & \text{if} \; t_k = t^e_n \; \text{is an exposure end},
\end{aligned}
\end{equation}
where
\begin{align}\label{eq:H_aug}
    \Haug_n = \frac{1}{\delta_n} \begin{pmatrix} 0 & 0 & 1\end{pmatrix}
\end{align}
picks out the integral state in $\xaug_k = [x, \dot{x}, z]^T$, which represents the integral from $t_n^s$ to $t_n^e$ (because we reset the integral state to zero at $t_n^s$) and divides by the exposure length $\delta_n$ to return the average of the state over that interval, which we measure with noise $\epsilon_n$. In other words, $y_n = z(t^e_n)/\delta_n + \epsilon_n$.

With Eqs.~\ref{eq:augmented_sde} and \ref{eq:augmented_observation_model} being of the form of Eqs.~\ref{eq:sde} and \ref{eq:measurement_model}, the augmented model with the integral state defines a state-space model which we can use the machinery of Section~\ref{sec:SSM} to solve. Next, we derive the Kalman and RTS equations to compute the predicted, filtered, and smoothed means and covariances at the $K = 2N$ states. These can then be used in Algorithm~\ref{alg:predict} as before to make predictions at arbitrary times.

\subsubsection{Integrated Kalman Filter}\label{sec:kalman_integrated}

Define the reset matrix $\RESET$ to be
\begin{align}
\RESET = \begin{pmatrix} 
            \I & 0 \\ 
            0 & 0
        \end{pmatrix}
\end{align}
so that the $\bm{x}$ state is preserved but the $z$ state is zeroed out. The only change to the Kalman filter is to apply the reset matrix as a transition matrix in the ``update'' step for exposure-start states. This can equivalently be thought of as a two-step transition to before and after the reset, where at each step we lack a measurement and so the update is skipped. We thus have (for exposure-start states),
\begin{align}
\text{Prediction:}& \nonumber \\
\m_{k}^{-} &= \Aaug_{k-1} \m_{k-1}, \nonumber \\
\P_{k}^{-} &= \Aaug_{k-1} \P_{k-1} \Aaug_{k-1}^T + \Qaug_{k-1}. \label{eq:integrated_kalman_pred} \\
\text{Update:}&   \nonumber \\
\m_k &= \RESET \m_{k}^{-}, \nonumber \\
\P_k &= \RESET \P_{k}^{-} \RESET^T. \label{eq:integrated_kalman_update}
\end{align}
In other words, the predicted exposure-start state is the same as the usual Kalman prediction from the previous state. We then treat the filtered exposure-start state as a deterministic ($\Q_k=0$) transition from that prediction, taking the reset matrix as our transition matrix. Exposure-end states are predicted and filtered according to the usual Kalman filter prescription (Eq.~\ref{eq:kalman_predict} and \ref{eq:kalman_update}) with the augmented matrices. 

This way, all the work to handle exposure integration is built-in to $z$ and $\Haug_n$; $z$ accumulates only during exposure intervals, at the end of which $\Haug_n$ projects the exposure-averaged value to the observed space.

\subsubsection{Integrated RTS Smoother}\label{sec:rts_integrated}

RTS smoothing for exposure-end states have the same form as Eq.~\ref{eq:rts}. For RTS smoothing at exposure-start states, we will have to smooth over both the usual $\Delta_k$ and also over the reset. As such, there are two changes to the usual RTS equations. The first is that, because $\hat{\m}_k,\,\hat{\P}_k$ are defined after smoothing over the reset, the corresponding quantities to smooth are $\m_k^-,\,\P_k^-$. The second is that, by interpreting the reset matrix as a transition matrix, we can think of forward transitions from ($\m_k^-,\,\P_k^-$) to their successive state ($\m_{k+1}^-,\,\P_{k+1}^-$) as a single transition where $\Aaug_k \RESET$ is the transition matrix. Substituting these two changes into Eq.~\ref{eq:rts}, we get the equations for smoothing an exposure-start state
\begin{equation}
\begin{aligned}
\label{eq:integrated_rts}
\G_{k} &= \P_k^- (\Aaug_k \RESET)^T\left[ \P_{k+1}^{-} \right]^{-1},  \\
\hat{\m}_k &= \m_k^- + \G_k \left[ \hat{\m}_{k+1} - \m_{k+1}^{-} \right], \\
\hat{\P}_{k} &= \P_k^- + \G_k \left[ \hat{\P}_{k+1} - \P_{k+1}^{-} \right] \G_k^T.
\end{aligned}
\end{equation}

Altogether, Eqs.~\ref{eq:integrated_kalman_pred}, \ref{eq:integrated_kalman_update}, and \ref{eq:integrated_rts} yield, respectively, the predicted ($\m^-_k$, $\P^-_k$), filtered ($\m_k$, $\P_k$), and smoothed ($\hat{\m}_k$, $\hat{\P}_k$) means and covariances at each of the $2N$ discretized states at all exposure starts and ends. The smoothed means and covariances at the exposure-end states, when projected through their corresponding $\Haug_n$, then give the conditioned mean ($\Haug_n \hat{\m}_n$) and variance ($\Haug_n\hat{\P}_n \Haug_n^T$) at each of the measurements. For making predictions at arbitrary times with Algorithm~\ref{alg:predict}, all $2N$ states must be used for accurate interpolation between (or during) measurements.

\subsubsection{Handling overlapping observations}\label{sec:overlap}

We assume, without loss of generality\footnote{Any set of overlapping measurements can be ``labeled'' as belonging to different instruments such that each unique instrument label does not contain any self-overlaps. The Kalman/RTS algorithms then scale as \bigO{N (d+N_\text{inst})^3}}, that observations from a single instrument do not overlap. That is, for each instrument $i \in [1,N_\text{inst}]$, $t^e_n < t^s_{n+1}\, \forall\; n \in N_i$. Then, we can introduce a $z_i$ state for each instrument into the augmented model so that our state becomes $\xaug = [\bm{x};z_1; z_2, \dots]^T$. The state dimension is now $d + N_\text{inst}$, and a new row and/or column is appended to each of the augmented matrices for each instrument,
\begin{equation}
\begin{aligned}
\Faug = 
\begin{pmatrix}
\F & 
  \begin{matrix}
    0\dots\\
    \vdots 
  \end{matrix}
\\[0.5em]
\begin{matrix}
  1 & 0\ldots \\
  \vdots & \vdots
\end{matrix}
&
  {}
\end{pmatrix}, 
&\quad \Laug = \begin{pmatrix} \L \\ 0 \\ \vdots \end{pmatrix}, \\
\Phiaug = \begin{pmatrix} 
                \bm{\Phi} & 0 & \dots  \\ 
                \Phibar_x & 1 &    \\
                \vdots &  & \ddots   \\
                \end{pmatrix},    
&\quad \Qaug = \begin{pmatrix} 
                \Q & \Qaug_{12}\dots   \\ 
                \Qaug_{21} & \Qaug_{22} \dots  \\
                \vdots & \vdots\quad\ddots  \\
                \end{pmatrix},   
\end{aligned}
\end{equation}
with zeros elsewhere. In other words, $\Faug$ has $\F$ as an upper-left block matrix, ones along the first column below that, and zeros elsewhere. $\Laug$ has $\L$ stacked above a zero for each instrument. $\Phiaug$ has $\bm{\Phi}$ as an upper-left block matrix, $\Phibar$ at the start of each row below that, ones along the diagonal, and zeros elsewhere. $\Qaug$ has $\Q$ as a block matrix in the upper-left, $\Qaug_{21}$ at the start of every row below that, $\Qaug_{12}$ repeatedly appended to the first column after $\Q$, and $\Qaug_{22}$ everywhere else. 

We then assign labels to each measurement start and end state to distinguish starts from ends, which measurement that start/end belongs to, and which instrument that measurement belongs to. Then, the only change to the Kalman filter is to reset each $z_i$ state \textit{only} at the start of an exposure belonging to that instrument. That way, $z_i$ is allowed to accumulate simultaneously with other instruments, thereby building in all cross correlations during overlaps by construction. Likewise, the RTS smoother must undo the reset for \textit{only} that instrument. That is, for instrument $i$ and integral state index $j$, the reset matrix is a block-diagonal matrix which is the identity except for the $z_j$ component matching the instrument,
\begin{align}
\RESET_i = \begin{pmatrix}
    \I & \\
       & 1-\delta_{i0} \\
       & & 1-\delta_{i1} \\
       & && \ddots
\end{pmatrix},
            \quad \delta_{ij} = \begin{cases}
            1 & i=j \\
            0 & i\neq j
            \end{cases}.
\end{align}
Then, exposure-end states in the Kalman filter project only that instrument's $z_i$.

Finally, the initial covariance $\tilde{\P}_0$ is undefined for the integral state. We choose to initialize to the identity for each integral state, though in practice it will not matter because these elements will be immediately reset upon reaching the first exposure-start state for that instrument. The identity is convenient because it makes $P_0$ invertible, which is needed for retrodiction (smoothing from the first exposure-start state),
\begin{align}
\P_0 = \begin{pmatrix}
    \Pinf & \\
       & 1 \\
       & & \ddots
\end{pmatrix}.
\end{align}

With these adjustments to the augmented model, and labels identifying the measurement and instrument at each exposure start and end state, we can simply iterate through the $2N$ states chronologically using the integrated Kalman and RTS prescriptions defined in the previous sections, resetting integral states at their corresponding instrument's exposure-start states and reading them off at the corresponding exposure-end states.

\begin{figure*}
    \centering
    \includegraphics[width=0.95\textwidth]{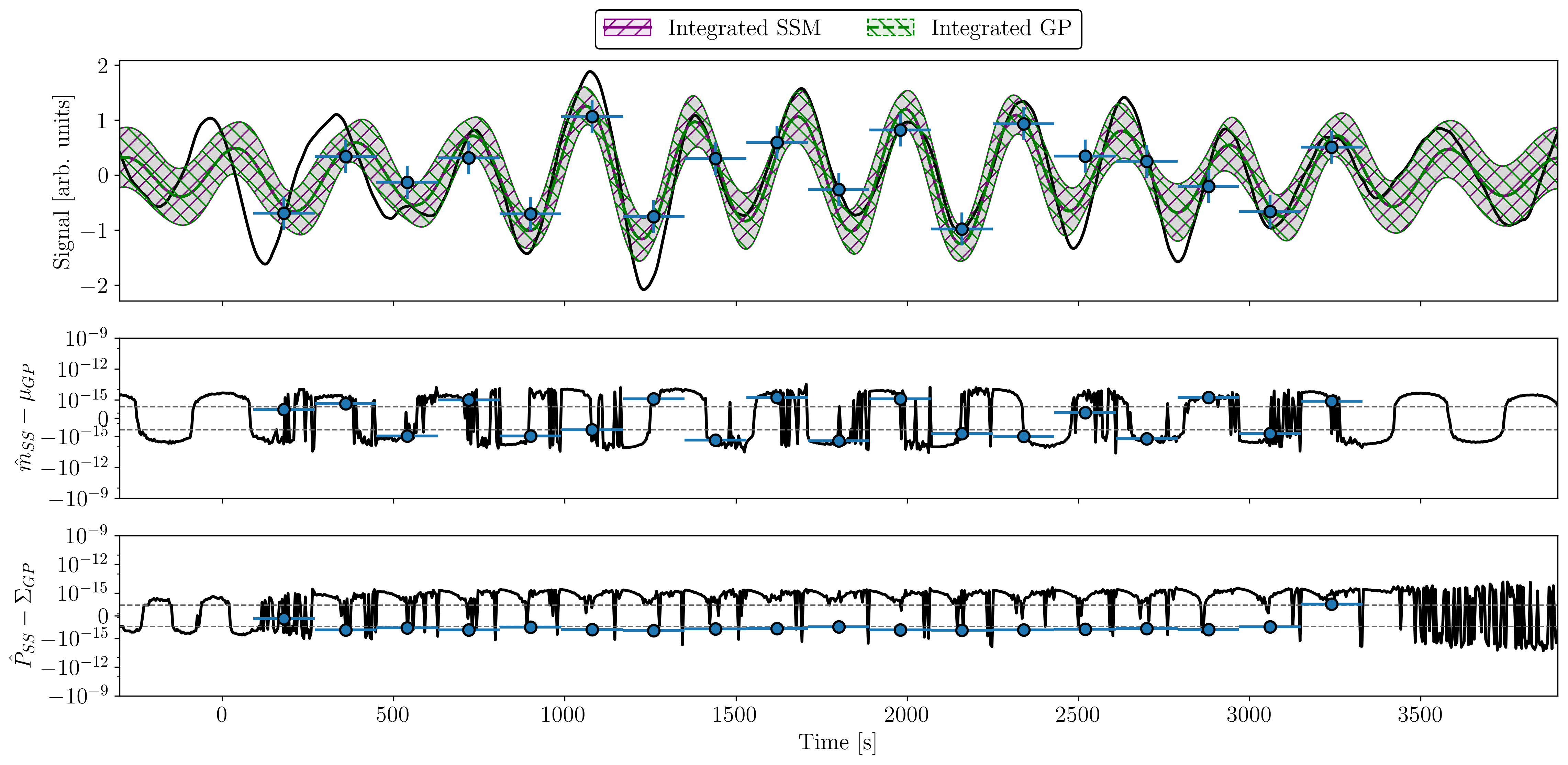}
    \caption{Numerical validation that the integrated SSM (purple curve) presented in this work produces the same result as the full integrated GP (green curve) from \citetalias{LuhnIntGP}. \textbf{Top:} An example stochastic signal (black curve) from an SHO kernel with a $\sim$300~sec timescale (Appendix~\ref{appendix:sho}). The data points are mock measurements of this curve with $180$~s exposures and uncertainties of 0.3 (arb. units). \textbf{Middle:} The difference between the conditioned (points) and predicted (black curve) mean of the SSM and GP approach. \textbf{Bottom:} The same for the variances. The horizontal dashed lines show machine epsilon for a single 64-bit floating-point numbers in the dynamic range of the simulated data; this also corresponds to the linear regime of the y-axis symmetric log scale. The two methods generally agree to within an order of magnitude of this level, consistent with the accumulated floating-point error over all the flops in the calculation. Periodicity in the residuals likely stem from the same periodicity in the kernel propagating through the various computations that yield the final mean and variance.}
    \label{fig:ss_vs_gp_integrated}
\end{figure*}

\begin{figure*}
    \centering
    \includegraphics[width=0.95\textwidth]{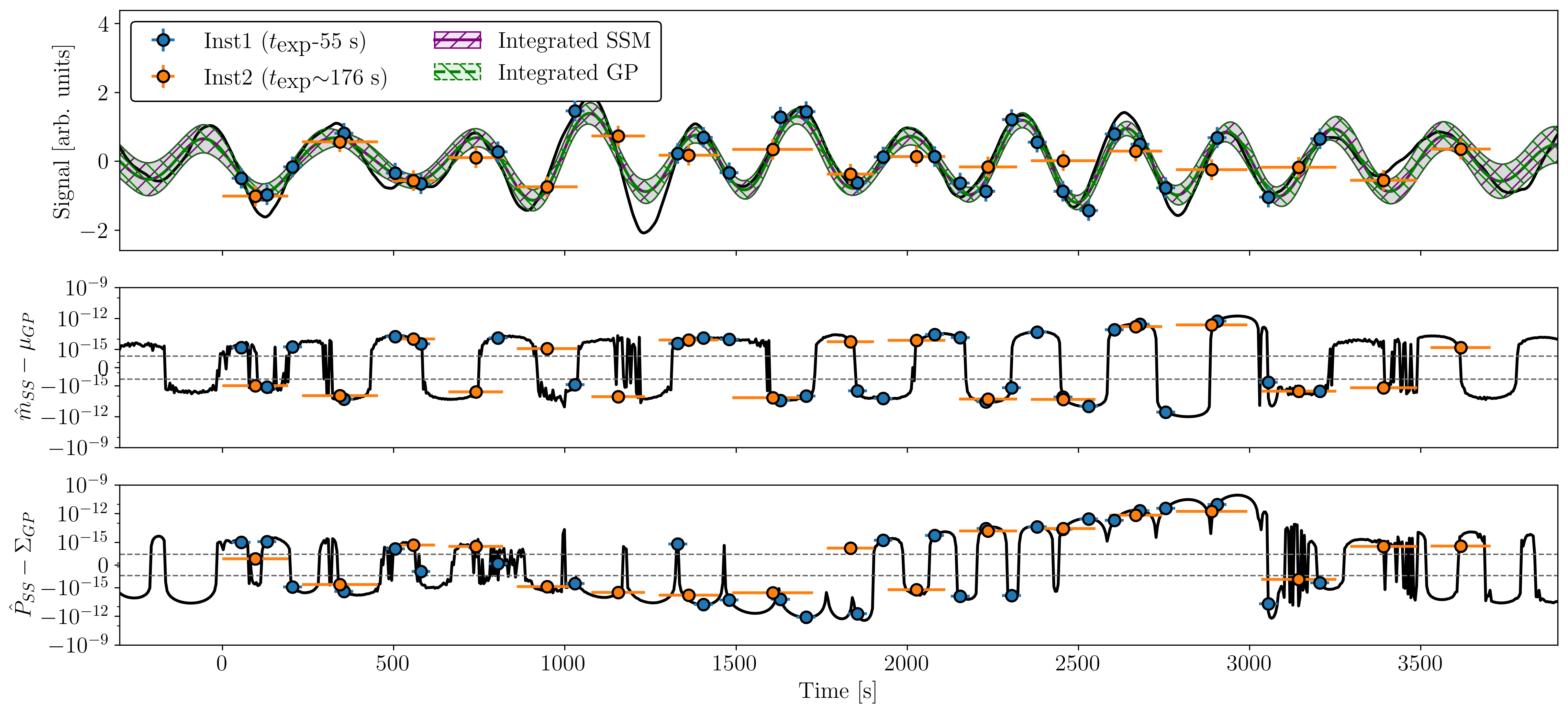}
    \caption{The same as Figure~\ref{fig:ss_vs_gp_integrated} but for a mock dataset with two instruments; one with 55~s exposures and one with variable $180 \pm 30$~s, which frequently overlap. Again, the two methods generally agree to within the expectation from numerical precision. Numerical stability in the variance may be improved with square-root filters \citep[e.g.,][]{Yaghoobi2025sqrt}.}
    \label{fig:ss_vs_gp_integrated_multiinstrument}
\end{figure*}

\subsubsection{Parallel Implementation}\label{sec:parallel}

The sequential Kalman and RTS algorithms discussed so far iterate once through the $K$ states in a forward pass and again in reverse. $K=N$ in the instantaneous case, and $K=2N$ in the integrated case; hence, the time complexity is \bigO{N}.

\citet{Sarkka2021} (hereafter \citetalias{Sarkka2021}) showed that, because of the associativity of the filtering ($p(x_k|y_{1:k})$) and smoothing ($p(x_k|y_{1:k},x_{k+1})$) distributions\footnote{And, consequently, the marginal likelihoods $p(y_{1:k})$.}, the sequential Kalman filter and RTS smoother can be reframed as an all-prefix-sums problem \citep[e.g.,][]{Blelloch1990}, which can be solved using a parallel-scan algorithm on a GPU. The parallel method reduces the wall-clock time complexity to \bigO{N/T + \log T}, for $T$ parallel workers. If $T \sim N$, the optimal scaling of \bigO{\log N} is achieved. For large datasets, usually $T \ll N$, giving \bigO{N/T}. An extensive experimental evaluation of these algorithms on GPUs can be found in \citet{Sarkka2025}, who also present a method to compute the parallel Kalman and RTS simultaneously on two GPUs. Nonlinear and square-root extensions of these methods can be found in \citet{Yaghoobi2021, Yaghoobi2025sqrt}.

Recently, \citet{Yaghoobi2025}, based on the original method of \citetalias{Sarkka2021}, implemented a parallel Kalman filter and RTS smoother for an integral observation model using fast and slow-rate states. In our case, we have the integral state augmented directly into the modeled state vector. As such, we can use the same formalism as \citetalias{Sarkka2021} to reframe Eq.~\ref{eq:integrated_kalman_update} and \ref{eq:integrated_rts} in terms of their associative parts. Here, we summarize the necessary changes to make the \citetalias{Sarkka2021} method compatible with integrated measurements.

Similar to the integrated sequential Kalman filter and RTS smoother, we handle the start and end states separately. For the parallel Kalman filter, the necessary associative parameters to calculate upfront for each of the $K$ steps are $(\bm{A}_k, \bm{b}_k, \bm{C}_k, \bm{\eta}_k, \bm{J}_k)$; note that these are defined as in \citetalias{Sarkka2021} (i.e., $\A_k$ is \textit{not} the transition matrix $\A_k$ as used previously in this manusript). The 5-tuple of associative parameters for a transition to an exposure-end state $k$ are given exactly by Eq.~10 of \citetalias{Sarkka2021}. For transitions to an exposure-start state, the reset requires us to set $\F_{\mathrm{eff}, k-1} = \RESET\, \F_{k-1}$ (in the associative notation, $\F_{k-1}$ is the transition matrix) and $\Q_{\mathrm{eff}, k-1} = \RESET\, \Q_{k-1}\, \RESET^T$. The lack of a measurement at $k$ requires us to set $R_k \leftarrow \infty$. The 5-tuple at an exposure-start state is then
\begin{equation} \label{eq:parallel_ikf_generic}
\begin{aligned} 
\bm{A}_k &= \F_{\mathrm{eff}, k-1}, \\ 
    \bm{b}_k &= \bm{0}, \\
    \bm{C}_k &= \Q_{\mathrm{eff}, k-1}, \\ 
    \bm{\eta}_k &= \bm{0}, \\
    \bm{J}_k &= \bm{0}.
\end{aligned}
\end{equation}
The initial $k=1$ step (Eq.~11 in \citetalias{Sarkka2021}) is also modified because it is always an exposure-start state for our problem formulation,
\begin{equation} \label{eq:parallel_ikf_first}
\begin{aligned} 
    \m^-_1 &= \F_{\mathrm{eff}, 0}\,\m_0, \\ 
    \P^-_1 &= \F_{\mathrm{eff}, 0}\,\P_0\,\F_{\mathrm{eff}, 0}^T + \Q_0, \\ 
    \bm{A}_1 &= \RESET, \\  
    \bm{b}_1 &= \m^-_1, \\
    \bm{C}_1 &= \P^-_1, \\
    \bm{\eta}_1 &= \bm{0}, \\
    \bm{J}_1 &= \bm{0}.
\end{aligned}
\end{equation}

For the parallel RTS smoother, we require a 3-tuple of associative parameters $(\bm{E}_k, \bm{g}_k, \L_k)$ for each of the $K$ steps. The 3-tuple at the final $k=K$ step (an exposure-end state) are unchanged from \citetalias{Sarkka2021}: 
\begin{equation}
\begin{aligned}
\label{eq:parallel_iRTS_final}
\bm{E}_K &= 0, \\
\bm{g}_K &= \m_K, \\
\bm{L}_K &= \P_K . 
\end{aligned}
\end{equation}

Likewise, the 3-tuple when $k$ is an exposure-end state is also the same as in \citetalias{Sarkka2021}. Exposure-start states are modified to match the behavior of Eq.~\ref{eq:integrated_rts}. Namely, they require the quantities to smooth be the predicted pre-reset state ($\m^-_k, \P^-_k$) and that $\F_{\mathrm{eff}, k} = \F_k\,\RESET$, giving
\begin{equation} \label{eq:parallel_iRTS_generic}
    \begin{aligned}
        \bm{E}_k &= \P^-_k\, \F_{\mathrm{eff}, k}^T \,(\F_{\mathrm{eff}, k}\,\P^-_k\,\F_{\mathrm{eff}, k}^T + \Q_k)^{-1}, \\
        \bm{g}_k &= \m^-_k - \bm{E}_k\,\F_{\mathrm{eff}, k}\,\m^-_k, \\
        \bm{L}_k &= \P^-_k - \bm{E}_k\,\F_{\mathrm{eff}, k}\,\P^-_k.
    \end{aligned}
\end{equation}

The $K$ 5-tuples for Kalman filtering (Eqs.~\ref{eq:parallel_ikf_generic} and \ref{eq:parallel_ikf_first} for exposure starts, Eq.~10 of \citetalias{Sarkka2021} for exposure ends) and the $K$ 3-tuples for RTS smoothing (Eqs.~\ref{eq:parallel_iRTS_final} and \ref{eq:parallel_iRTS_generic} for exposure starts, \citetalias{Sarkka2021} for exposure ends) are each combined via the appropriate binary associative operator \citepalias[see][]{Sarkka2021}. We implemented this method in the  \texttt{ParallelIntegratedStateSpaceSolver} class of \smolgp\ using the \texttt{JAX} parallel scan function \texttt{jax.lax.associative\_scan}.

\section{Validation and Performance}\label{sec:benchmark}

First, we numerically validated the equivalence between our integrated SSM and the integrated GP framework of \citetalias{LuhnIntGP}. For the former, we used the \texttt{IntegratedStateSpaceModel} class of \smolgp, and for the latter we defined a custom \tinygp\ kernel which implements the various cases and (sub)integrals defined in \citetalias{LuhnIntGP} for all pairs of integrated measurements. We generated a single and multi-instrument synthetic dataset by sampling from a ``true'' latent signal, defined by an SHO GP. We then computed the conditioned mean and variance at the data points (integrated over the exposure interval) as well as the predicted latent (instantaneous) curve at a dense grid of test points. 

Figure~\ref{fig:ss_vs_gp_integrated} shows a representative example with a single instrument for the edge case of no deadtime between exposures. Both the integrated SSM and GP methods agree at the level of machine precision when working with 64-bit floating-point numbers. 

A multi-instrument example is shown in Figure~\ref{fig:ss_vs_gp_integrated_multiinstrument} which has instrument 1 taking constant 55~s exposures \citep[similar to NEID solar data,][]{Lin2022} and instrument 2 taking variable exposure times sampled from a Gaussian with mean $180$~s and standard deviation $30$~s \citep[similar to EXPRES solar data,][]{LOST}. In this case, the two methods generally agreed at the level expected from numerical precision when working with 64-bit floating-point numbers. Likewise, the likelihoods computed by all methods agreed to within $10^{-14}.$ We also verified that the parallel method gives the same result as the sequential solver in all cases. 

\begin{figure*}
\centering
\includegraphics[width=0.32\textwidth]{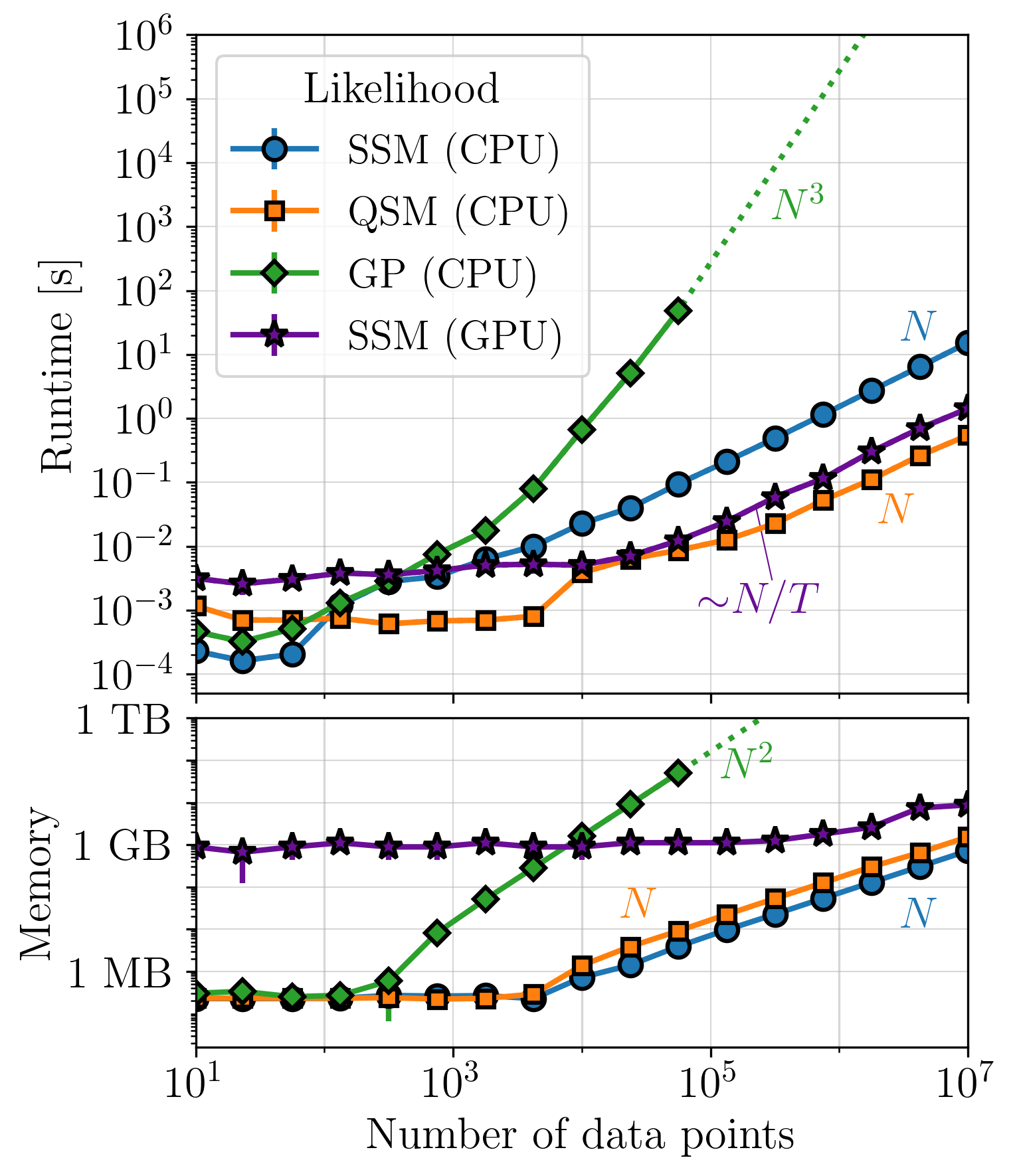}
\includegraphics[width=0.32\textwidth]{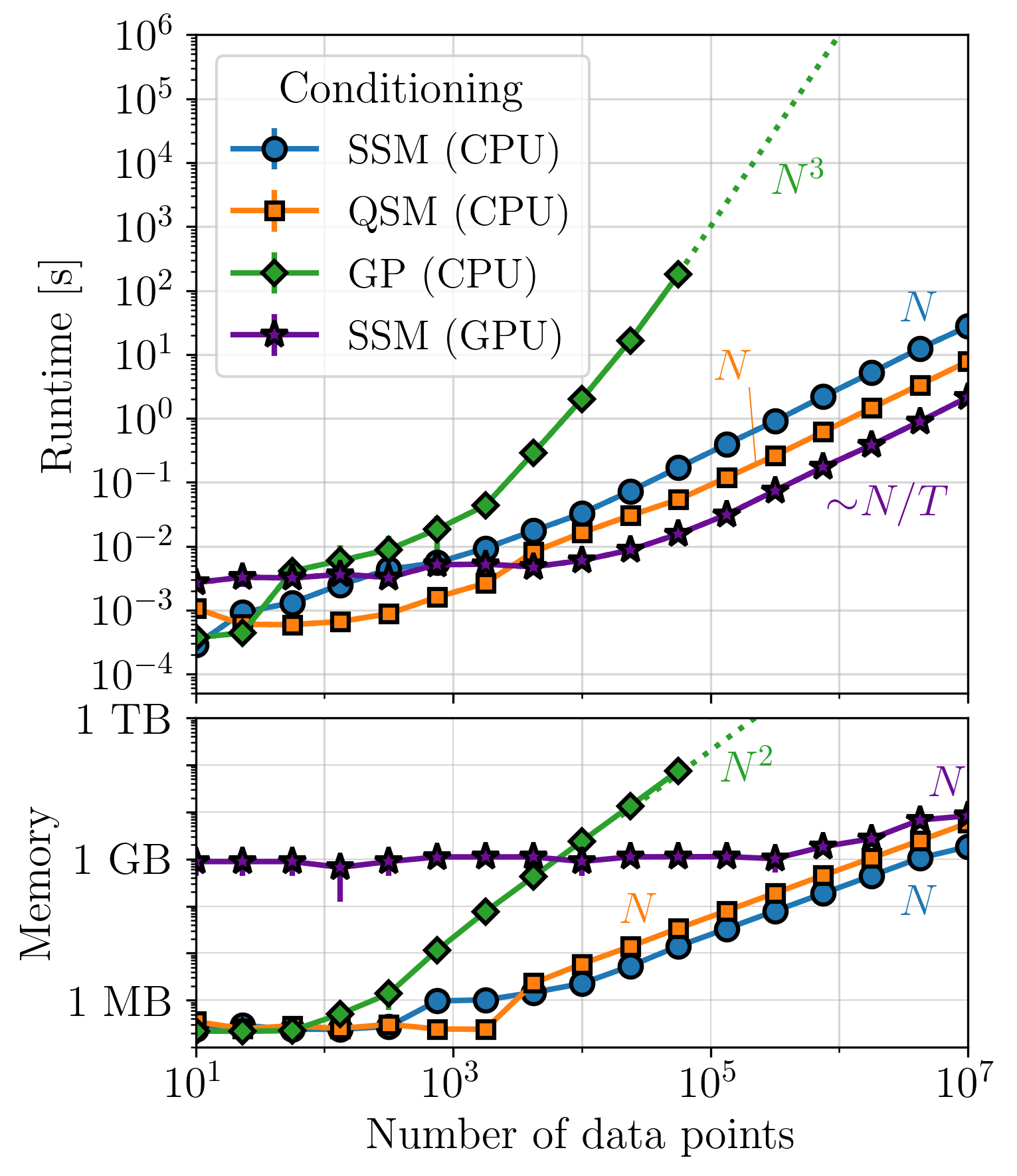}
\includegraphics[width=0.32\textwidth]{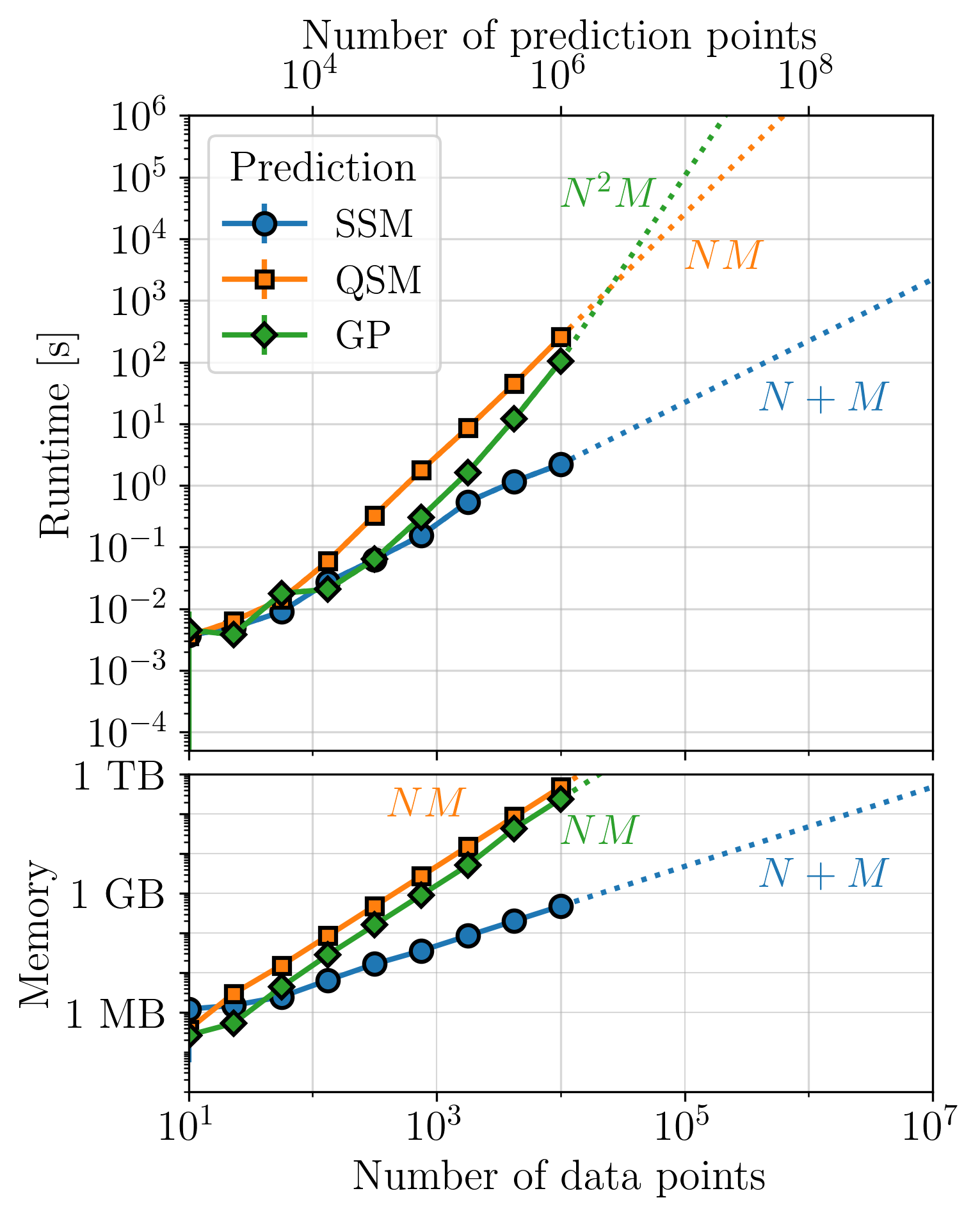}
    \caption{Runtime and memory benchmarking comparing the performance on instantaneous data of the full/dense GP solution (i.e. Eq.~\ref{eq:full GP solution}, as in \tinygp, green diamonds), using QSM algebra (also via \tinygp, orange squares), and our implementation of the sequential SSM solver (\smolgp, blue circles); these three cases were all tested on a CPU because their performance on a GPU was degraded. Dashed lines trace the theoretical scaling from the largest value tested. In all cases, the top panel shows the wall-clock timed average of five runs, while the bottom shows the peak memory usage during the function execution. The purple stars show the parallel SSM solver (\citetalias{Sarkka2021} as implemented in \smolgp) as tested on a NVIDIA RTX 6000 Ada GPU running CUDA v12.8. \textbf{Left:} Results for the log-likelihood as a function of $N$ data points. \textbf{Middle:} Results for conditioning at the $N$ data points, including initialization. \textbf{Right:} Results for conditioning on $N$ data points and then predicting at $M = 100N$ test points, to simulate a typical high-resolution prediction scenario. \textbf{Takeaway:} The SSM shares the linear runtime scaling as QSMs but is typically more memory efficient (especially for predictions). Computing the likelihood is faster in the QSM framework, although for conditioning, the best runtime performance is achieved by the parallel SSM (with high memory overhead on a GPU).}
    \label{fig:benchmark}
\end{figure*}

\begin{figure*}
\centering
\includegraphics[width=0.32\textwidth]{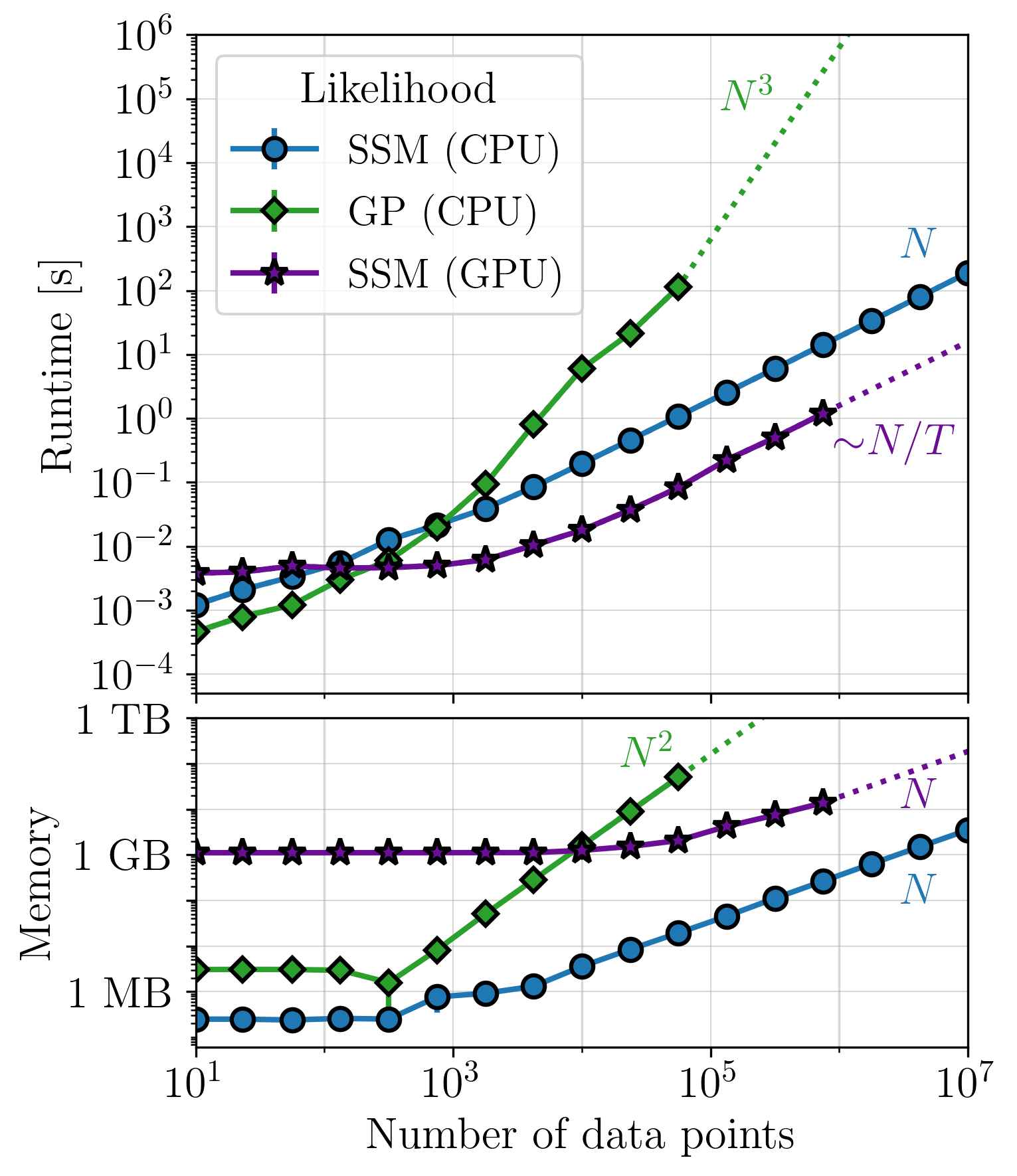}
\includegraphics[width=0.32\textwidth]{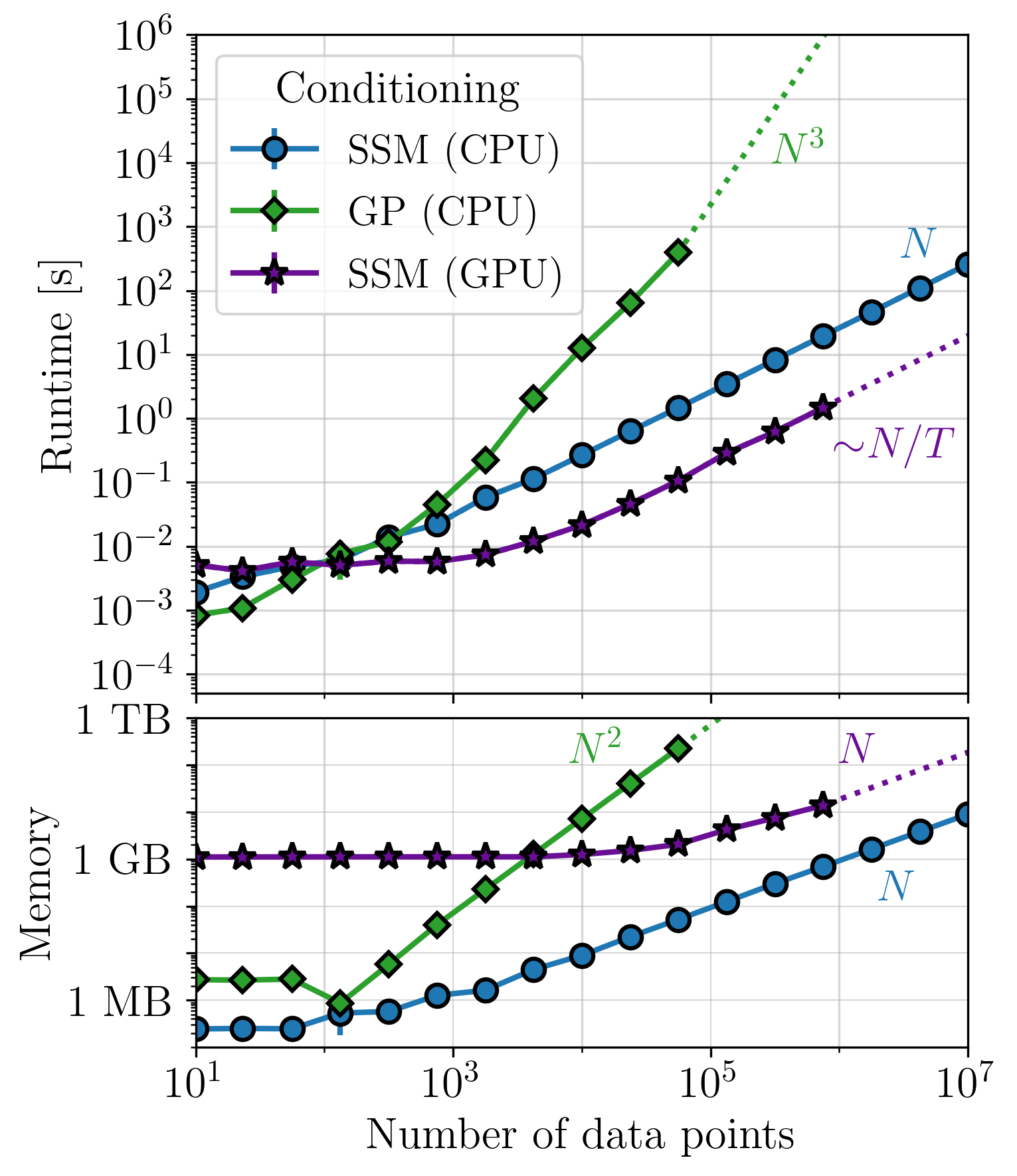}
\includegraphics[width=0.32\textwidth]{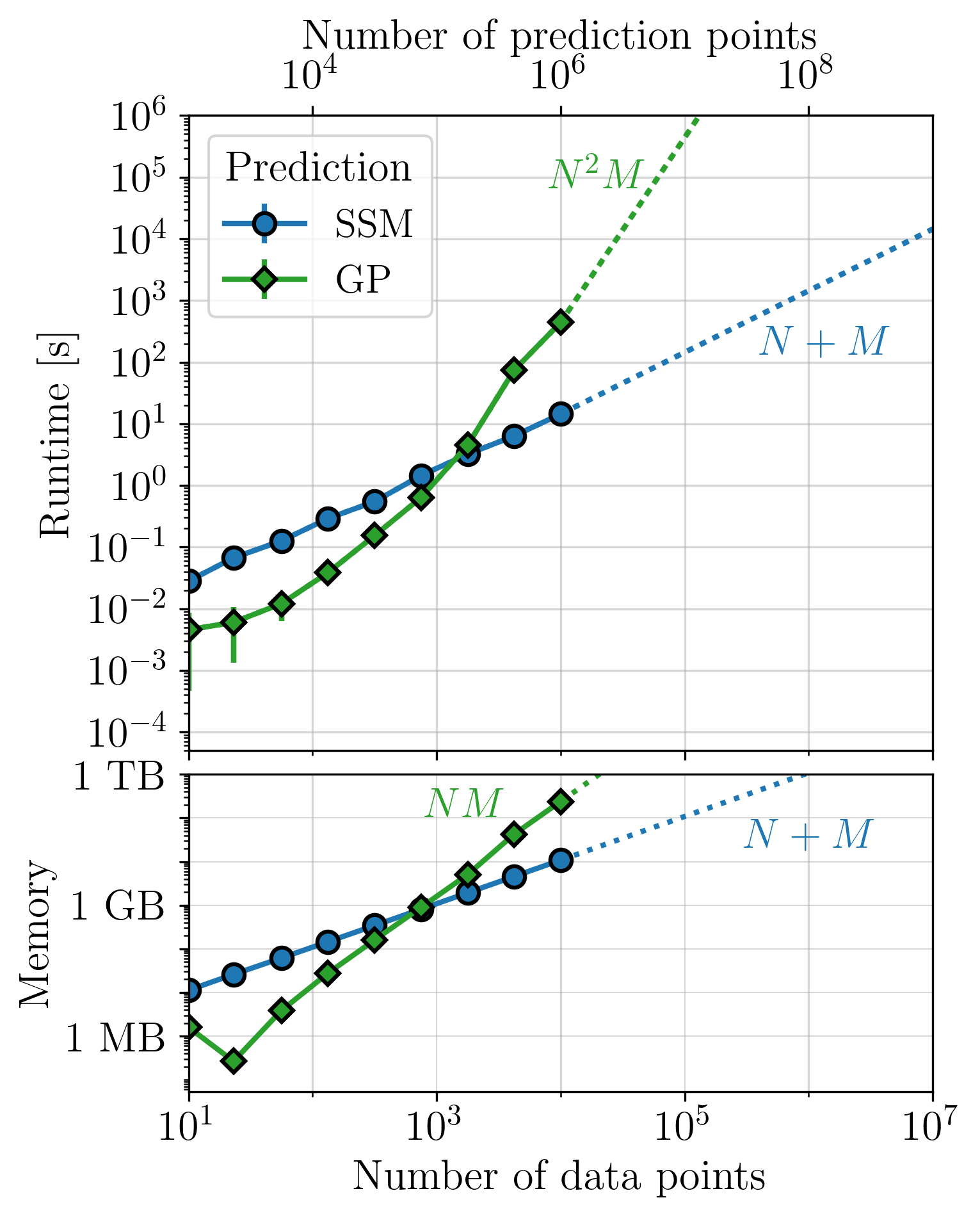}
    \caption{Same as Figure~\ref{fig:benchmark} but with integrated measurements. The full/dense GP solution (green diamonds) uses the \citetalias{LuhnIntGP} method, which we implemented in \tinygp. The SSM solution (blue circles) uses our augmented SSM approach (Section~\ref{sec:integrated_ssm}), as implemented in \smolgp. Likewise, the parallel version (purple circles) implements Section~\ref{sec:parallel} in \smolgp. \textbf{Takeaway:} Because there is no quasi-separable framework compatible with integrated measurements, the SSM method provides a transformative speedup and memory savings over traditional GP methods which are forced to construct the full covariance matrix.}
    \label{fig:benchmark_integrated}
\end{figure*}

Finally, we benchmarked the full GP, QSM, and SSM methodologies to compare performance. We only show the performance for the full GP, QSM, and sequential SSM solvers as tested on a CPU (Intel$^{\text{\textregistered}}$
 Xeon$^{\text{\textregistered}}$ w5-3435X) because their performance was degraded on a GPU; conversely, we only show the parallel SSM solver as tested on a GPU (NVIDIA RTX 6000 Ada, CUDA v12.8) because its CPU performance was degraded. Our setup was as follows. For each function to benchmark (likelihood, conditioning, and prediction), we set up \tinygp\ objects for the full and QSM representations of an SHO kernel and a \smolgp\ object for the SSM representation. We took a random draw from the prior distribution as our true signal, from which we generated random measurements of varying size $N$. We timed five computations of each function at each $N$, while monitoring their peak memory usage in a separate thread; this peak memory use does not include overheads before calling the function (e.g. $\sim$200~MB from importing \texttt{JAX} and JIT compilation).

The average and standard deviation of the five runs, for both the runtime and the peak memory use, are shown in Figure~\ref{fig:benchmark} for the tests involving instantaneous data and Figure~\ref{fig:benchmark_integrated} for those involving integrated data. The most dramatic improvements are seen with integrated data, since there is no QSM representation to compete with the SSM method. Generally, we found the SSM method to be the most memory efficient, which can easily become limiting at modest $N \sim 10^5$ with the full GP method ($\gtrsim$100~GB of RAM), as well as for predictions with the QSM method. As tested, the SSM method is the fastest for making predictions, and when parallelized on a GPU is the fastest at conditioning and rivals QSM methods at computing the likelihood.

\section{Summary and Discussion}\label{sec:conclusion}

We have developed a state-space framework for solving the general GP regression problem when the data contain integrated, possibly overlapping, measurements. Compared to the method of \citetalias{LuhnIntGP} which constructs the full, dense covariance matrix and integrates the covariance function between all pairs of data points (accounting for overlap), our method yields equivalent results but can be solved in linear (and even logarithmic) time complexity. This is achieved by augmenting a linear Gaussian SSM with an integral state $z$ such that $dz/dt = x$, which we reset to zero at the start of the exposures and make noisy measurements of (normalized by the exposure length) at the ends of the exposures. 

We derived the matching Kalman filter and RTS smoother (Section~\ref{sec:integrated_ssm}) and demonstrated numerical equivalence to \citetalias{LuhnIntGP} (Section~\ref{sec:benchmark}) to near machine precision, though lingering structured residuals likely stem from some numerical instabilities in the filtering/smoothing algorithms; these may be remedied with square-root filters, which can also be parallelized \citep{Yaghoobi2025sqrt}. We also presented a speed and memory optimal method, Algorithm~\ref{alg:predict}, for predicting at arbitrary points once the conditioned means and covariances at all the states have been obtained. Overall, the SSM framework, especially its compatibility with parallel methods (Section~\ref{sec:parallel}), brings integrated and multi-instrument datasets into the same (or faster) regime as state-of-the-art QSM methods, opening the door to holistic analyses of massive multi-instrument datasets ($N > 10^5$). 

The integrated SSM-GP approach discussed here is not just limited to time series, but any binned data where the independent coordinate is sortable (i.e. there must be a unique causal sequence for the Kalman/RTS scans). Likewise, multivariate data (i.e. parallel time series) are also compatible with an appropriately shaped observation matrix; the example used to demonstrate the parallel method in \citetalias{Sarkka2021} is a 2D trajectory through space. SSMs can also handle spatiotemporal models \citep{SarkkaHartikainen2012,Sarkka2013}, nonlinear models \citep{nonlinear}, non-Gaussian/non-conjugate likelihoods \citep{nongaussian}, certain nonstationary kernels \citep{nonstationary, Lin2025}, model learning problems \citep{modellearning}, and even time-variable hyperparameters \citep[via inducing points,][]{Liu2021}. These flexible extensions of SSMs could provide significant utility to other, more complex astronomical modeling problems which demand scalable methods. Additionally, because the modeled latent state contains not just the instantaneous state but also its derivatives, SSMs are a natural, powerful choice for modeling phenomena that include derivative observations; one example is the FF' model \citep{Aigrain2012, Rajpaul2015, Jones2017, Gilbertson2020b, Tran2023}.

We implemented the full SSM-GP framework discussed throughout this manuscript in the Python/\texttt{JAX} package \smolgp, which is available (under the MIT license) at \href{https://github.com/smolgp-dev/smolgp}{https://github.com/smolgp-dev/smolgp}. The \smolgp\ design philosophy is to have as identical an API as possible to \tinygp\ \citep{tinygp} so existing analyses can easily incorporate these new features.  We encourage contributions and engagement from the community on how best to expand these methods to other astronomical problems of interest.

\begin{acknowledgments}

The authors gratefully thank the anonymous referee for their thoughtful feedback which greatly improved clarity throughout the manuscript. We also thank Lehman Garrison, Joseph Long, David Hogg, Adrian Price-Whelan, and the CCA Astro Data and Software groups for engaging and helpful discussions on methodology, code design, algorithms, and statistics. 
R.A.R. was supported by the Flatiron Research Fellowship at the Flatiron Institute, a division of the Simons Foundation.

\end{acknowledgments}

\software{
\texttt{smolgp}   \citep{smolgp},
\texttt{tinygp}   \citep{tinygp},
\texttt{JAX}      \citep{jax},
\texttt{matplotlib} \citep{matplotlib}
}

\bibliography{references}
\bibliographystyle{aasjournal}


\appendix

\section{Exposure time integration for celerite models} \label{sec:celerite_integration}

Here, we give expressions for the exposure time integrated versions of the \celerite\ kernel and its power spectrum.  It turns out that the time-integrated \celerite\ kernel remains expressible as a semiseparable matrix when the exposures do not overlap, but when exposures overlap for nonzero time lags, the kernel no longer is semiseparable.

For the \celerite\ model, the general kernel function is expressed as \citep{celerite}
\begin{equation}
k(\Delta) = a e^{-c\tau}\cos(d\Delta) + b e^{-c\tau}\sin(d\Delta),
\end{equation}
where $\Delta = \vert t_i-t_j \vert$ is the lag between times $t_i$ and $t_j$, and the parameters of the model are $a$, $b$, $c$, and $d$.
If the two measurements have exposure time $\delta$, then the kernel becomes the following integral \citepalias[using the notation of][]{LuhnIntGP},
\begin{align}
k_{FF}(\delta, \Delta) = \frac{1}{\delta^2}\int_{t_i-\frac{\delta}{2}}^{t_i+\frac{\delta}{2}} \int_{t_j-\frac{\delta}{2}}^{t_j+\frac{\delta}{2}} k(|t - t'|) \dd t \dd t'.
\end{align}

When $\delta \le \Delta$, the exposures do not overlap and the integrated model still falls within the class of \celerite\ models
\begin{equation}\label{eq:delta-small}
k_{FF}(\delta, \Delta|\, \delta \leq \Delta) = A e^{-c \Delta} \cos{(d \Delta)} + B e^{-c \Delta} \sin{(d \Delta)},
\end{equation}
where $c$ and $d$ have not changed, but the amplitudes are given by the updated expressions
\begin{equation}
  A = \frac{2C_1\left[\cosh(c \delta) \cos(d \delta) - 1\right]
  - 2 C_2 \sinh(c \delta) \sin(d \delta)
}{\delta^2 (c^2 + d^2)^2},
\end{equation}
and
\begin{equation}
  B = \frac{2 C_2 \left[\cosh(c \delta) \cos(d \delta) - 1\right]
  + 2 C_1 \sinh(c \delta) \sin(d \delta)}{\delta^2 (c^2 + d^2)^2},
\end{equation}
with
\begin{align}
C_1 = a c^2 - a d^2 + 2 b c d
\quad\mathrm{and}\quad
C_2 = b c^2 - b d^2 - 2 a c d.
\end{align}

For the case of overlapping exposures ($0 \le \Delta < \delta$), the result is not as simple because of the absolute value in the definition of $\Delta$. In this case, we take $t_i > t_j$ without loss of generality, and separate the integral into three integrals that can be easily evaluated
\begin{align}
k_{FF}(\delta, \Delta|\,\Delta < \delta)
&= \frac{1}{\delta^2 } \left( \int_{t_j+\frac{\delta}{2}}^{t_i+\frac{\delta}{2}}  \int_{t_j-\frac{\delta}{2}}^{t_j+\frac{\delta}{2}} k(t - t^\prime) \dd t' \dd t
+ \int_{t_i-\frac{\delta}{2}}^{t_j+\frac{\delta}{2}} \int_{t_j-\frac{\delta}{2}}^{t} k(t - t^\prime) \dd t' \dd t
+ \int_{t_i-\frac{\delta}{2}}^{t_j+\frac{\delta}{2}} \int_{t}^{t_j+\frac{\delta}{2}} k(t^\prime - t) \dd t' \dd t \right).
\end{align}
Evaluating and simplifying the result gives
\begin{align}
k_{FF}(\delta, \Delta|\, \Delta < \delta) =& \; k_{FF}(\delta, \Delta|\, \delta\le\Delta) + 2 \left[
  (a c + b d) (c^2 + d^2) (\delta-\Delta) \right. \nonumber\\
&  - C_1 \sinh(c \delta-c \Delta) \cos(d \delta-d \Delta) \nonumber \\
&  \left. + C_2 \cosh(c \delta-c \Delta) \sin(d \delta-d \Delta) \right] / [\delta (c^2+d^2)]^2,
\end{align}
where the first term is the result from Eq.~\ref{eq:delta-small}. One can see that this cannot be expressed as a \celerite\ term.

The power spectrum of the time-integrated term derived above can be computed by taking the Fourier transform and, after some simplifications, it reduces to the expected result
\begin{align}
S(\delta, \omega) &= S(\omega) \mathrm{sinc}^{2}(\omega \delta/2),
\end{align}
where $S(\omega)$ is the power spectrum of the original term and $\mathrm{sinc}(x) = \sin(x)/x$ is the Fourier transform of a top hat representing the exposure. 
An interesting consequence of this form is that, since the sinc function goes to zero when $\omega \delta = 2 \pi n$ for integers $n \ne 0$, there is no power at integer multiples of the sampling frequency $f = \frac{2\pi}{\omega} = \delta^{-1}$.

An important implementation note is that, while a nonintegrated \celerite\ term would generate a matrix with the amplitude $a$ on the diagonal, the diagonal for the integrated term is not simply $A$ because the lag on the diagonal is $\Delta = 0 < \delta$.
This means that the diagonal entries in the covariance matrix for a time-integrated \celerite\ model are
\begin{align}
k_{FF}(\delta, \Delta=0) = A + \frac{2 \left[ (a c + b d) (c^2 + d^2) \delta 
  - C_1 \sinh(c \delta) \cos(d \delta) 
  + C_2 \cosh(c \delta) \sin(d \delta)\right]}{\delta^2 (c^2+d^2)^2}
\end{align}
which can be simplified to
\begin{equation}
k_{FF}(\delta, \Delta=0) = \frac{2 \left[ (a c + b d) (c^2+d^2) \delta - C_1 + e^{-c \delta} \left(C_1 \cos(d \delta)+C_2 \sin(d \delta)\right) \right]}{\delta^2 (c^2 + d^2)^2}.
\end{equation}

For the case of $b=0$ and $c=0$ (called the ``real'' term by \citealt{celerite}), the integrated kernel simplifies to
\begin{equation}
  k_{FF}(\delta,\Delta) = \frac{2 a e^{-c \Delta}}{\left(c \delta\right)^2} 
    \begin{cases}
    e^{-c \delta} - 1 + c \delta & \Delta = 0 \\
    \left[c (\delta - \Delta) - \sinh{(c \delta-c \Delta)}\right] e^{c \Delta} + \cosh{(c \delta)} -1 & 0 < \Delta \le \delta \\
    \cosh(c \delta) - 1  & \delta < \Delta
    \end{cases} \quad.
\end{equation}
In \celerite, this exposure-averaging treatment is handled by the \texttt{TermConvolution} term.

\section{SSM for the Stochastically Driven Damped Harmonic Oscillator}\label{appendix:sho}

In this appendix, we work out the full state-space representation of the SHO, starting from the GP kernel. While the SHO is our kernel of interest that requires exposure-integrated treatment (to model asteroseismic oscillations), it is also an excellent pedagogical choice since it has a familiar SDE that will guide us to the state-space form. The general steps are: kernel function $\rightarrow$ PSD $\rightarrow$ transfer function $\rightarrow$ SDE $\rightarrow$ state-space matrices.


The GP kernel for an SHO with parameters $S_0$, $\omega_0$, and $Q$ has the following form \citep{celerite},
\begin{align}\label{eq:sho_kernel}
k(\Delta) &= \sigma^2 \exp\left(-\frac{\omega_0}{2Q}\Delta\right)
        \begin{cases}
            1 + \omega_0 \Delta & \text{for } Q = 1/2, \\
            \cos(\tau)  + \frac{1}{2\eta Q}\sin(\tau)  & \text{for } Q > 1/2, \\
            \cosh(\tau) + \frac{1}{2\eta Q}\sinh(\tau) & \text{for } Q < 1/2, 
        \end{cases} \tag{Kernel}
\end{align}
where we have used the shorthand variables $\sigma = \sqrt{S_0\omega_0Q}$, $\tau = \eta\omega_0\Delta$, and $\eta = \sqrt{|1 - 1/(4Q^2)|}$. Per the Wiener-Khintchine theorem, the corresponding PSD is
\begin{align}\label{eq:sho_psd}
S(\omega) = \frac{2 S_0\,\omega_0^4}
        {(\omega^2-{\omega_0}^2)^2 + {\omega_0}^2\,\omega^2/Q^2}, \tag{PSD}
\end{align}
which is clearly a rational function with all the dependence on $\omega$ in the denominator. The next step, called \textit{minimum-phase spectral factorization} (e.g., Chapter 9 of \citealt{optimalfiltering}) in fields such as signal processing, is to identify the transfer function $H(i \omega)$ that satisfies Eq.~\ref{eq:transfer} and defines a causal and stable (i.e. minimum phase) system. 
We can notice that the denominator of Eq.~\ref{eq:sho_psd} can be factored like so:
\begin{align}
    S(\omega) = \frac{2 S_0 \omega_0^4}{\omega^4 - 2\omega_0^2\omega^2 + \frac{\omega_0^2}{Q^2}\omega^2 + \omega_0^4} = \frac{2 S_0 \omega_0^4}{\left((i\omega)^2 + \frac{\omega_0}{Q}(i\omega) + \omega_0^2\right)\left((-i\omega)^2 + \frac{\omega_0}{Q}(-i\omega) + \omega_0^2\right)}.
\end{align}
Once we define 
\begin{align}\label{eq:sho_q}
    q = 2 S_0 \omega_0^4 = \frac{2\omega_0^3}{Q} \sigma^2,
\end{align}
we are left with a complex conjugate pair of possible $H(i \omega)$ that both satisfy Eq.~\ref{eq:transfer},
\begin{align}
    H_+(i \omega) = \frac{1}{(i\omega)^2 + \frac{\omega_0}{Q}(i\omega) + \omega_0^2}, \quad
    H_-(i \omega) = \frac{1}{(-i\omega)^2 + \frac{\omega_0}{Q}(-i\omega) + \omega_0^2}.
\end{align}
The minimum-phase option is identified by extending these two functions to the complex $s$-plane and selecting the one whose poles lie in the left half-plane, i.e., whose poles satisfy $\mathrm{Re}(s) < 0$. That is, the appropriate transfer function for the SHO is
\begin{align}\label{eq:sho_transfer}
    H(s) = H_+(s)= \frac{1}{s^2 + \frac{\omega_0}{Q}s + \omega_0^2}. \tag{Transfer function}
\end{align}
Since the transfer function of an LTI system is, by construction, the ratio of the output to the input in the frequency domain, with the assumption that all initial conditions are zero (e.g., \citealt{eerb}, \citealt{linear-system-theory}), we can use Eq.~\ref{eq:ratio} to get the SDE in the Laplace domain,
\begin{align}\label{eq:sho_laplace_sde}
    H(s) = \frac{X(s)}{W(s)} \; \rightarrow \; s^2 X(s) + s\frac{\omega_0}{Q}X(s) + \omega_0^2 X(s) = W(s),
\end{align}
where $X(s)$ and $W(s)$ are the Laplace transforms of $x(t)$ and $w(t)$ (the output and the input, respectively).
Inverse Laplace transforming Eq.~\ref{eq:sho_laplace_sde}, again under the assumption that all initial conditions are zero, yields the SDE for the latent state $x(t)$ of an SHO,
\begin{align}\label{eq:sho_sde}
\frac{\dd^2 x}{\dd t^2} + \frac{\omega_0}{Q} \frac{\dd x}{\dd t} + \omega_0^2 x(t) = w(t), \tag{SDE}
\end{align}
which we can recognize from physics. To define the SSM, we want to put this into the form of Eq.~\ref{eq:sde},
\begin{align}
\frac{\dd\bm{x}}{\dd t} = \F\bm{x}(t) + \L w(t),
\end{align}
which we can do by setting $\bm{x}(t) = [x, \dot{x}]^T$ and reading off the coefficients to build $\F$ and $\L$ (Eq.~\ref{eq:generic_F}):
\begin{align}
    \F = \begin{pmatrix}
                0       &       1 \\
            -\omega_0^2 & -\frac{\omega_0}{Q}
         \end{pmatrix},
         \qquad
\L = \begin{pmatrix}
            0 \\ 1
          \end{pmatrix}.
\end{align} 

To get $Q_c$ and $\Pinf$, we use these $\F$ and $\L$ and define a dummy $\Pinf = \begin{pmatrix}P_{11} & P_{12} \\ P_{21} & P_{22} \end{pmatrix}$ and substitute into Eq.~\ref{eq:lyapunov_eq}:
\begin{align}
\begin{pmatrix} 0 & 1 \\ -\omega_0^2 & -\frac{\omega_0}{Q} \end{pmatrix}
\begin{pmatrix}P_{11} & P_{12} \\ P_{21} & P_{22} \end{pmatrix} + 
\begin{pmatrix}P_{11} & P_{12} \\ P_{21} & P_{22} \end{pmatrix} 
\begin{pmatrix} 0 & -\omega_0^2\\ 1 & -\frac{\omega_0}{Q} \end{pmatrix} + 
\begin{pmatrix} 0 & 0 \\ 0 & Q_c \end{pmatrix} = 0. \nonumber
\end{align}
Working out the multiplications yields four equations which we can solve for each element of $\Pinf$,
\begin{align}
\Pinf = \begin{pmatrix} \frac{Q Q_c}{2\omega_0^3} & 0 \\ 0 & \frac{Q Q_c}{2\omega_0} \end{pmatrix}.
\end{align}
Then, setting the $P_{\infty,11}$ element (stationary variance in the state) equal to $k(\Delta=0) = \sigma^2$ we have

\begin{align} \label{eq:sho_Qc_Pinf}
Q_c = \frac{2\omega_0^3}{Q}\sigma^2,  \qquad
\bm{P}_\infty = \sigma^2 \begin{pmatrix}
            1 & 0 \\ 
            0 & \omega_0^2
            \end{pmatrix}.
\end{align} 
Note $Q_c$ is exactly $q$ in Eq.~\ref{eq:sho_q}, as expected.

\citet{Jordan2021} worked out $\A_k$ and $\Q_k$ for the Mat{\'e}rn family and SHO. They derived full analytic expressions for $\exp(\F\Delta)$ and the Lyapunov integral for $\Q_k$ using a Laplace transform, matrix factorization, and a symbolic mathematics package. A perhaps easier way is to use the well-known solution $x(t)$ to the SDE for a given kernel and read off $\A_k$ from its definition
\begin{align}
\bm{x}(t) &= \exp\left(\F t\right) \bm{x}(t=0) \nonumber \\
          &=\A_k \bm{x}(t=0) \nonumber \\
\rightarrow \; \A_k &= \bm{x}(t) [\bm{x}(t=0)]^{-1},
\end{align}
and then determine $\Q_k$ from Eq.~\ref{eq:Q} using $\A_k$ and $\Pinf$.

In any case, the result for the SHO is
\begin{align}\label{eq:sho_phi}
    \A_{k} = \Phi(\Delta_k) = 
    \begin{cases}
        e^{-\omega_0\Delta_k}
            \begin{pmatrix}
                    1 + \omega_0\Delta_k  & \Delta_k \\
                    -\omega_0^2 \Delta_k  & 1 - \omega_0\Delta_k
            \end{pmatrix} & \text{for } Q = 1/2 \\
            \\
        e^{-\frac{\omega_0\Delta_k}{2Q}}
            \begin{pmatrix}
                    \cosarg + \frac{1}{2\eta Q}\sinarg   &   \frac{1}{\eta\omega_0} \sinarg  \\
                    \frac{-\omega_0}{\eta}\sinarg       &   \cosarg - \frac{1}{2\eta Q}\sinarg
            \end{pmatrix} & \text{for } Q > 1/2 \\
            \\
        e^{-\frac{\omega_0\Delta_k}{2Q} }
            \begin{pmatrix}
                    \cosharg + \frac{1}{2\eta Q}\sinharg   &   \frac{1}{\omega_0\eta} \sinharg  \\
                    \frac{-\omega_0}{\eta} \sinharg   &   \cosharg - \frac{1}{2\eta Q}\sinharg
            \end{pmatrix} & \text{for } Q < 1/2
    \end{cases}
\end{align}

where we again used the shorthand $\shoarg = \eta\omega_0\Delta_k$. We can then derive $\Q_k$ by substituting the above $\A_k$ into Eq.~\ref{eq:Q} given $\Pinf$ (Eq.~\ref{eq:sho_Qc_Pinf}). We obtained (for the underdamped $Q > 1/2$)
\begin{align}\label{eq:sho_Q}
    \Q_{k} = \sigma^2 e^{-\frac{\omega_0\Delta_k}{Q}}
            \begin{pmatrix}
                    \Qaa  &  \Qab  \\
                    \; \\
                    \Qba  &  \Qbb
            \end{pmatrix},
\end{align}
We note that Eq.~\ref{eq:sho_Q} differs from that in \citet{Jordan2021} (their Eq. 26) by a factor of $Q/2\omega_0^3 \sigma^2$ on the diagonals and $Q/\sigma^2$ on the off-diagonals. In fact, their expression omits $S_0$ entirely. We verified Eq.~\ref{eq:sho_Q} is correct by both numerically integrating Eq.~\ref{eq:Q} as well as computing $\Q$ from a Van Loan matrix exponential (e.g. Eq.~\ref{eq:Q_from_vanloan})--all were equivalent. Importantly, using this definition in an augmented SSM with an integral state yielded identical results to the full integrated SHO kernel of \citetalias{LuhnIntGP} (see Figure~\ref{fig:ss_vs_gp_integrated}). 

With Eq.~\ref{eq:sho_phi} in hand, we can compute the integrated transition matrix,
\begin{align}\label{eq:sho_phibar}
    \Phibar(\Delta) &= \int_0^\Delta \bm{\Phi}(t) \dd t \nonumber \\
            &= \frac{e^{at}}{a^2+b^2}\left.
               \begin{pmatrix}
               \Ic + A (\Is) & B (\Is) \\
               C (\Is) & \Ic - A (\Is)
               \end{pmatrix}\right|_{t=0}^{t=\Delta}
\end{align}
where here $\tau = \eta\omega_0 t$, and $A = 1/2\eta Q$, $B = 1/\eta\omega_0$, $C=-\omega_0/\eta$, $a = -\omega0/2Q$, and $b=\eta\omega_0$.

For datasets where the numerical value of $\F \Delta$ is never larger than $\sim$$10^5$, it is computationally fast and efficient to obtain $\Qaug$ from $\Faug$ and $\Laug Q_c \Laug^T$ via its Van Loan matrix exponential \citep{VanLoan1978}
\begin{align}\label{eq:Q_from_vanloan}
    \Qaug(\Delta) = \F_3^T \bm{G}_2, \;\; \text{where} \; \exp\left[ \begin{pmatrix}
                -\Faug  &  \Laug Q_c \Laug^T \\
                \bm{0} & \Faug^T
                \end{pmatrix}  \Delta\right] &= \begin{pmatrix}
                \F_2 & \bm{G}_2 \\ \bm{0} & \F_3
                \end{pmatrix}.
\end{align}
However, since we have $\Q$ analytically (Eq.~\ref{eq:sho_Q}), a more numerically stable means of constructing $\Qaug$ is to assemble it in its block form (Eq.~\ref{eq:Q_aug}). The top-left block is just $\Q$. The remaining blocks can be computed numerically from Eqs. 1.3 ($\bm{M}(\Delta)$) and 1.4 ($\bm{W}(\Delta)$) in \citet{VanLoan1978},
\begin{align}
\Qaug(\Delta) = \begin{pmatrix}
\Q(\Delta) & \bm{M}(\Delta) \\ \bm{M}(\Delta)^T & \bm{W}(\Delta)
\end{pmatrix} = \begin{pmatrix}
                \Q(\Delta) & \bm{F}_3^T \bm{H}_2 \\
                (\bm{F}_3^T \bm{H}_2)^T & (\bm{F}_3^T \bm{K}_1) + (\bm{F}_3^T \bm{K}_1)^T
                \end{pmatrix},
\end{align}
where
\begin{align}
\exp\left[ \begin{pmatrix}
            -\F & \I  & \Z   & \Z \\
            \Z  & -\F & \L Q_c \L^T  & \Z \\
            \Z  & \Z  & \F^T & \I \\
            \Z  & \Z  & \Z   & \Z \\   
            \end{pmatrix} \Delta
    \right] = 
    \begin{pmatrix}
        \bm{F}_1 & \bm{G}_1 & \bm{H}_1 & \bm{K}_1 \\
        \Z & \bm{F}_2 & \bm{G}_2 & \bm{H}_2 \\
        \Z & \Z & \bm{F}_3 & \bm{G}_3& \\
        \Z & \Z & \Z & \bm{F}_4 \\
    \end{pmatrix}.
\end{align}
Recall that because we only use the first row of $\Phibar$ in $\Phiaug$, we likewise only need the first column of $\bm{M}$, which we call $\tilde{\Q}_{12}$, the first row of $\bm{M}^T$ (which is just $\tilde{\Q}_{12}^T$), and the top-left element of $\bm{W}$, which we call $\tilde{Q}_{22}$, to give $\Qaug$ the correct shape
\begin{align}\label{eq:analytic_Qaug}
    \Qaug(\Delta) = \begin{pmatrix}
                \Q(\Delta)  &  \Qaug_{12}(\Delta) \\
                \Qaug_{12}^T(\Delta) & \tilde{Q}_{22}(\Delta)
                \end{pmatrix}.
\end{align}
In the case of the SHO, we can instead derive these analytically from 
\begin{align}
 \Qaug_{12} &\equiv \text{first column of}\; \int_0^\Delta \bm{\Phi}(t) \L \Qc \L^T \Phibar(t)^T \dd t, \\
\Qaug_{22} &\equiv\text{first element of}\; \int_0^\Delta \Phibar(t) \L \Qc \L^T \Phibar(t)^T \dd t.
\end{align}

For $Q > 1/2$, these are
\begin{align}\label{eq:analytic_Qaug12}
    \Qaug_{12}(\Delta) = \sigma^2\begin{pmatrix}
        \frac{1}{Q \omega_0}[\exparg (\cosarg + A \sinarg) - 1]^2 \\
        A\exparg[4\sinarg - \exparg \sin(2\shoarg)] - 2A^2\expaarg\sin^2(\shoarg) + \expaarg-1
        \end{pmatrix}, 
\end{align}
and
\begin{gather}
\begin{aligned}\label{eq:analytic_Qaug22}
\tilde{Q}_{22}(\Delta) =& \frac{\sigma^2}{4Q^2\omega_0^2}\left[
    8Q\omega_0\Delta + 4Q^2 - 12 +A^2 \expaarg (\cos(2\shoarg) - 16Q^4) \right. \\
    &\left.\qquad\qquad + 16\exparg (\cosarg + (1 - 2Q^2)A\sinarg)
    + \expaarg\left(\frac{1 - 3A^2}{A}\sin(2\shoarg) - 3\cos(2\shoarg)\right) \right],
\end{aligned}
\end{gather}
where again $\tau = \eta\omega_0 \Delta$ and $A = 1/2\eta Q$. We can see from Eq.~\ref{eq:analytic_Qaug22} that $\tilde{Q}_{22}$ grows linearly with $\Delta$ as $\Delta \rightarrow \infty$, which is the source of numerical instability when computing this term via matrix exponentials.

\section{Multicomponent Kernels}\label{appendix:multicomponent}

\subsection{Sums of kernels}\label{appendix:sums}

In the GP framework, a sum of covariance kernels is itself a GP kernel \citep{gpml}. That is,
\begin{align}
\bm{K} = \bm{K}_1 + \bm{K}_2 + \dots + \bm{K}_M = \sum_{m=1}^{M} \bm{K}_m 
\end{align}

also defines a GP. In the state-space framework, a model which is the sum of other SSMs can analogously be constructed by stacking the components in a block-diagonal form \citep[Algorithm 12.9 in][]{SarkkaSDEbook}

\begin{align}
\F &= \text{blkdiag}(\F_1, \F_2, \dots),      &       \bm{L}   &= \text{blkdiag}(\L_1, \L_2, \dots),  \\
\Qc &= \text{blkdiag}(\Q_{c,1}, \Q_{c,2}, \dots),      &       \Pinf &= \text{blkdiag}(\P_{\infty,1}, \P_{\infty,2}, \dots), \nonumber
\end{align}

The matrix exponential $\bm{\Phi}(\Delta) = \exp(\F \Delta)$ is also simply the block-diagonal matrix of component matrix exponentials, per the definition of the matrix exponential
\begin{align}
\bm{\Phi}(\Delta) &= \exp(\text{blkdiag}(\F_1, \F_2, \dots) \Delta) \nonumber \\
                  &= \text{blkdiag}(\bm{\Phi}_1, \bm{\Phi}_2, \dots)(\Delta).
\end{align}
Likewise the process noise also takes on a block-diagonal form:
\begin{align}
    \Q &= \text{blkdiag}(\Q_1, \Q_2, \dots).
\end{align}
The rest of the machinery (Kalman/RTS) proceeds as before. Each component is therefore treated independently in the process dynamics but becomes coupled by the stacked measurement model $\H = (\H_1 \;\H_2\; \dots)$. The overall state dimension is the sum of the component models, $d = \sum_{m=1}^M d_m$.

\subsection{Products of kernels}\label{appendix:products}

Likewise, products of kernels also produce valid covariance kernels \citep{gpml}. That is,
\begin{align}
\K = \K_1 \K_2 \dots \K_M = \prod_{m=1}^{M} \K_m
\end{align}
defines a GP. Algorithm 12.10 in \citep{SarkkaSDEbook} gives the machinery for the product of SSMs,
\begin{gather}
\F = \F_1 \otimes \I + \I \otimes \F_2 \dots,   \\
\begin{aligned}
\Pinf &= {\Pinf}_1 \otimes {\Pinf}_2 \dots, & \H &= \H_1 \otimes \H_2 \dots \nonumber
\end{aligned}
\end{gather}

where $\otimes$ is the Kronecker product (which also yields block matrices). As a result, the dimension of model is now the product of the component dimensions, $d = \prod_{m=1}^M d_m$, making products much more expensive than sums of kernels in the state-space.

The matrix exponential $\A = \bm{\Phi}(\Delta) = \exp(\F \Delta)$ also becomes the Kronecker product of the component matrix exponentials,
\begin{align}
\bm{\Phi}(\Delta) &= \exp(\F \Delta) = \exp(\F_1 \Delta \otimes \I + \F_2 \Delta \otimes \I \dots) \nonumber \\
                  &= \bm{\Phi}_1(\Delta) \otimes \bm{\Phi}_2(\Delta) \dots
\end{align}

Substituting the above $\F$ and $\Pinf$ into the Lyapunov equation (Eq.~\ref{eq:lyapunov_eq}), for a product of two kernels, we get
\begin{align}\label{eq:product_LQL}
    \L \Qc \L^T = (\L_1 \Q_{c,1} \L_1^T) \otimes \P_{\infty,2} + \P_{\infty,1} \otimes (\L_2 \Q_{c,2} \L_2^T).
\end{align}
Any $\L$ and $\Qc$ that satisfy this are valid; a convenient choice is $\L = \I$ and $\Qc$ equal to the right-hand side of Eq.~\ref{eq:product_LQL}. Relatedly, $\Q$ is best determined from $\Pinf$ and $\A$ via Eq.~\ref{eq:Q}, which can be factored for a product of two kernels as
\begin{align}\label{eq:product_Q}
\Q = \P_{\infty,1}\otimes\Q_2 + \Q_1\otimes\P_{\infty,2} - \Q_1\otimes\Q_2.
\end{align}

\subsection{Component-wise predictive means and variances}\label{appendix:component_means}

Say the kernel is a sum of $M$ component kernels,

\begin{align}
\bm{K} = \bm{K}_1 + \bm{K}_2 + ... + \bm{K}_M = \sum_{m=1}^{M} \bm{K}_m.
\end{align}

In GP regression, we can compute the predictive mean and variance from any of the individual kernel components by substituting that component kernel into Eq.~\ref{eq:full GP solution} like so

\begin{align}
\bm{\mu}_{GP,m} &= \bm{K}_{m, \ast}^T (\bm{K} + \bm{N})^{-1} \bm{y}     \nonumber \\
\bm{\Sigma}_{GP,m} &= \bm{K}_{m,\ast\ast} - \bm{K}_{m,\ast}^T (\bm{K} + \bm{N})^{-1} \bm{K}_{m,\ast}.  
\end{align}

Basically, wherever the kernel is computed with test points, we use the component kernel of interest. The overall mean and variance can be reconstructed from the components via  \citep{Duvenaud2014}

\begin{align}
\bm{\mu}_{GP} = \sum_{m=1}^M \bm{\mu}_{GP,m}, \qquad \bm{\Sigma}_{GP} = \sum_{m=1}^M \bm{\Sigma}_{GP,m} - \sum_{m\neq l}^M \bm{K}_{m,\ast}^T\bm{K}^{-1}\bm{K}_{l,\ast}.  
\end{align}

To do the same in the state-space framework, recall that the structure of a sum/product of kernels given in the previous sections treat the dynamics of each component as independent (i.e., the matrices are combined in block-diagonal form). Thus, the Kalman/RTS algorithms by construction provide us with all of the component means and variances, which get summed together when projecting through the coupled observation matrix $\H = (\H_1 \;\H_2\; \dots)$. To pick out just one component, we can simply do
\begin{equation}
\begin{aligned}
\bm{\mu}_{GP,m} &= \bm{H}^m \hat{\m}, \\
\bm{\Sigma}_{GP,m} &= \bm{H}^m \hat{\P} {\bm{H}^m}^T 
\end{aligned}
\end{equation}
where
\begin{equation}
\begin{aligned}
\H^m  = (0,\dots,\H_m,0,\dots) \\
\end{aligned}
\end{equation}
picks out just the observation matrix for the component of interest. 

\end{document}